\documentclass[sigplan,10pt]{acmart}

\renewcommand\footnotetextcopyrightpermission[1]{} 
\acmConference{}{}{}
\acmYear{}
\acmISBN{} 
\acmDOI{} 

\setcopyright{none}

\usepackage{amsmath}
\usepackage{booktabs}   
\usepackage{subcaption} 
\usepackage{paralist}
\usepackage{cleveref}
\usepackage{graphicx,color}
\usepackage{listing}
\usepackage{listings}
\usepackage{mathpartir}
\usepackage{mdframed}
\usepackage{tikz}
\usetikzlibrary{arrows, fit, positioning, shapes}
\usepackage{xcolor}
\usepackage{xspace}
\usepackage{mathtools}
\usepackage[subtle]{savetrees}

\newcommand{\Libra}{$\textsc{Anon}$ }

\NewEnviron{centerframe}[1][]{%
    \begin{mdframed}
		\begin{center}
		\BODY
		\end{center}
		\end{mdframed}
}

\newcommand{\yz}[1]{}

\newcommand{\commentout}[1]{}

\newcommand{\tup}[1]{\left<#1\right>\xspace}

\newcommand{\set}[1]{\left\{#1\right\}\xspace}
\newcommand{\brackets}[1]{\left[#1\right]\xspace}
\newcommand{\dom}[1]{dom({#1})\xspace}
\newcommand{\img}[1]{img({#1})\xspace}

\newcommand{\move}{Move\xspace}

\newcommand{\redacted}[1]{URL hidden for double-blind review.}

\newcommand{\cl}{\ensuremath{\mathit{Loc}}}

\newcommand{\val}{\ensuremath{Val}}
\newcommand{\primval}{\ensuremath{\mathit{PrimVal}}}
\newcommand{\primtype}{\ensuremath{\mathit{PrimType}}}
\newcommand{\valtype}{\ensuremath{\mathit{ValType}}}
\newcommand{\reftype}{\ensuremath{\mathit{RefType}}}
\newcommand{\type}{\mathit{Type}}
\newcommand{\mutreftype}{\mathit{MutRefType}}
\newcommand{\immreftype}{\mathit{ImmRefType}}
\newcommand{\imm}[1]{\mathit{Imm}(#1)}
\newcommand{\mut}[1]{\mathit{Mut}(#1)}

\newcommand{\fn}{\ensuremath{F}}

\newcommand{\record}[1]{\set{#1}\xspace}
\newcommand{\typeof}[2]{{#1}:{#2}}
\newcommand{\cons}{{::}}
\newcommand{\stackht}[2]{{#1}\cons{#2}\xspace}

\newcommand{\rff}{\ensuremath{{\bf ref}\!\!\!}}  

\newcommand{\rft}[2]{\rff~\tup{{#1}, {#2}}\xspace}

\newcommand{\ite}[3]{\mathit{ite}(#1, #2, #3)}

\newcommand{\st}[1]{\tup{#1}\xspace}

\newcommand{\ao}[2]{\ensuremath{{#1}\tup{#2}}\xspace}

\newcommand{\mdel}[2]{{#1}\setminus{#2}}
\newcommand{\mset}[3]{{#1}\brackets{{#2} \mapsto {#3}}}
\newcommand{\mread}[2]{#1(#2)\xspace}

\newcommand{\lset}[3]{{#1}\brackets{{#2} \mapsto {#3}}}
\newcommand{\lread}[2]{#1(#2)\xspace}

\newcommand{\cmdname}[1]{{\ensuremath{\bf {#1}}}\xspace}
\newcommand{\rulename}[1]{\text{\bf {#1}}\xspace}

\newcommand{\program}{\ensuremath{\mathcal{P}}\xspace}
\newcommand{\proc}{\ensuremath{\rho}}

\newcommand{\movelocCmd}{\cmdname{MoveLoc}}

\newcommand{\copylocCmd}{\cmdname{CopyLoc}}

\newcommand{\storelocCmd}{\cmdname{StoreLoc}}

\newcommand{\borrowlocCmd}{\cmdname{BorrowLoc}}

\newcommand{\readrefCmd}{\cmdname{ReadRef}}

\newcommand{\writerefCmd}{\cmdname{WriteRef}}

\newcommand{\freezerefCmd}{\cmdname{FreezeRef}}

\newcommand{\callCmd}{\cmdname{Call}}

\newcommand{\returnCmd}{\cmdname{Ret}}

\newcommand{\packCmd}{\cmdname{Pack}}

\newcommand{\unpackCmd}{\cmdname{Unpack}}

\newcommand{\borrowfieldCmd}{\cmdname{BorrowField}}

\newcommand{\movetoCmd}{\cmdname{MoveTo}}

\newcommand{\movefromCmd}{\cmdname{MoveFrom}}

\newcommand{\borrowglobalCmd}{\cmdname{BorrowGlobal}}

\newcommand{\popCmd}{\cmdname{Pop}}

\newcommand{\stackopCmd}{\cmdname{Op}}

\newcommand{\stackop}{\cmdname{Op}}

\newcommand{\absL}{\ensuremath{\mathit{ALoc}}}

\newcommand{\branchCmd}{   \cmdname{Branch}}

\newcommand{\op}{\ensuremath{{\mathrm op}}}
\newcommand{\pc}{\ensuremath{\ell}}

\newcommand{\pcstep}[3]{\ensuremath{{#1}\vdash{#2}\rightarrow{#3}\xspace}}
\newcommand{\pcrule}[5][]{\ensuremath{\inferrule*[Right=\rulename{#1}]{#2}{\pcstep{#3}{#4}{#5}}}\xspace}

\newcommand{\lstates}[1]{\program[#1].\mathit{T}}
\newcommand{\astates}[1]{\program[#1].\overline{\mathit{T}}}
\newcommand{\bcodes}[1]{\program[#1].\mathit{C}}
\newcommand{\ins}[1]{\program[#1].\mathit{I}}
\newcommand{\outs}[1]{\program[#1].\mathit{O}}
\newcommand{\progloc}[2]{\bcodes{#1}\brackets{#2}}

\newcommand{\sep}{\quad}

\newcommand{\elim}[2]{\mathit{elim}(#1,#2)}
\newcommand{\rename}[2]{\mathit{rename}(#1,#2)}
\newcommand{\factor}[3]{\mathit{factor}(#1,#2,#3)}
\newcommand{\factorf}[4]{\mathit{factor}_{#1}(#2,#3,#4)}
\newcommand{\extend}[3]{\mathit{extend}(#1,#2,#3)}
\newcommand{\is}{\mathit{is}}
\newcommand{\os}{\mathit{os}}
\newcommand{\reverse}[1]{\mathit{rev}(#1)}

\newcommand{\abs}[1]{\mathit{Abs}(#1)}
\newcommand{\absframe}[1]{\mathit{AP}(#1)}
\newcommand{\absstack}[1]{\mathit{AS}(#1)}
\newcommand{\absgraph}[2]{\mathit{AB}(#1,#2)}

\newcommand{\matchcmd}{\mathit{match}\xspace}
\newcommand{\letcmd}{\mathit{let}\xspace}
\newcommand{\incmd}{\mathit{in}\xspace}
\newcommand{\ifcmd}{\mathit{if}\xspace}
\newcommand{\thencmd}{\mathit{then}\xspace}
\newcommand{\elsecmd}{\mathit{else}\xspace}

\newcommand{\Next}[1]{\mathit{Next}(#1)}

\newcommand{\booleantype}{{\tt Bool}\xspace}
\newcommand{\integertype}{{\tt Int}\xspace}
\newcommand{\addresstype}{{\tt Addr}\xspace}

\newcommand{\concState}{s}
\newcommand{\ls}{\mathit{ls}}

\newcommand{\Borrow}{\mathit{Borrow}}

\newcommand{\absState}{\hat{\concState}}
\newcommand{\psloc}{\Pi}
\newcommand{\osloc}{\Omega}
\newcommand{\len}[1]{\mathit{len}(#1)}
\newcommand{\below}{\sqsubseteq}

\newcommand{\inv}{\mathit{Inv}}
\newcommand{\concat}[2]{#1::#2}

\newcommand{\inva}{\mathit{InvA}}
\newcommand{\invb}{\mathit{InvB}}
\newcommand{\invc}{\mathit{InvC}}
\newcommand{\invd}{\mathit{InvD}}

\newcommand{\us}{\mathit{us}}
\newcommand{\vs}{\mathit{vs}}

\newtheorem{manualtheoreminner}{Theorem}
\newenvironment{manualtheorem}[1]{%
  \renewcommand\themanualtheoreminner{#1}%
  \manualtheoreminner
}{\endmanualtheoreminner}

\definecolor{pblue}{rgb}{0.13,0.13,1}
\definecolor{pgreen}{rgb}{0,0.5,0}
\definecolor{pred}{rgb}{0.9,0,0}
\definecolor{pgrey}{rgb}{0.46,0.45,0.48}

\definecolor{dullred}{Hsb}{0,1,0.4}
\definecolor{dullyellow}{Hsb}{30,1,0.4}
\definecolor{dullgreen}{Hsb}{60,1,0.4}
\definecolor{dullteal}{Hsb}{150,1,0.4}
\definecolor{dullblue}{Hsb}{210,1,0.4}
\definecolor{dullpurple}{Hsb}{270,1,0.4}
\definecolor{dullmagenta}{Hsb}{300,1,0.4}
\definecolor{dullpurplered}{Hsb}{330,1,0.4}

\definecolor{ckeyword}{HTML}{7F0055}
\definecolor{ccomment}{HTML}{3F7F5F}
\definecolor{cnumber}{HTML}{2A0099}

\lstdefinelanguage{Move}{
 keywords={
   abort, acquires, assert, copy, borrow_global, borrow_global_mut, create_account, freeze, fun,
   module, move_to, move_from, public, resource, break, exists,
    continue,
    else,
    false,
    if,
    let, loop,
    move, mut,
    return,
    struct,
    true,
    while,
    use,
  },
  ndkeywords={address, bool, u64, bytearray, Self},
  showspaces=false,
  showtabs=false,
  breaklines=true,
  showstringspaces=false,
  breakatwhitespace=true,
  lineskip=-0.6pt,
  morecomment=[l]{//}, 
  morecomment=[s]{/*}{*/}, 
  basewidth={0.54em, 0.4em},%
  basicstyle=\footnotesize\ttfamily,
  keywordstyle={\color{ckeyword}\ttfamily\bfseries},
  ndkeywordstyle={\color{pblue}\ttfamily\bfseries},
  commentstyle={\color{ccomment}\itshape},
  stringstyle=\color{green},
  moredelim=[il][\textcolor{pgrey}]{$$},
  moredelim=[is][\textcolor{pgrey}]{\%\%}{\%\%}
}

\lstdefinestyle{default}{%
  basicstyle=\ttfamily,%
  commentstyle=\sl,%
  keywordstyle=\bf,%
  columns=fullflexible,%
  keepspaces=true,%
  mathescape,%
  escapechar=\#%
}
\lstdefinestyle{number}{%
  numbers=left,%
  numberstyle=\scriptsize\em,%
}
\newcommand{\mcode}[1]{\lstinline[language=Move,basicstyle=\small\ttfamily]{#1}}
\newcommand{\code}[1]{\mcode{#1}}

\newcommand{\borrow}[3]{\color{dullteal}#1 \xRightarrow{#2} #3}
\newcommand{\reject}{\color{dullteal}\dagger}
\newcommand{\emptygraph}{\color{dullteal}\emptyset}

\begin{document}
\pagestyle{plain}

\title{The Move Borrow Checker}


\author{Sam Blackshear*}
\affiliation{
  \institution{Mysten Labs}
  \country{USA}
}
\email{sam@mystenlabs.com}          

\author{John Mitchell}
\affiliation{
  \institution{Stanford}
  \country{USA}
}
\email{jcm@stanford.edu}          

\author{Todd Nowacki}
\affiliation{
  \institution{Mysten Labs}
  \country{USA}
}
\email{todd@mystenlabs.com}          

\author{Shaz Qadeer}
\affiliation{
  \institution{Meta}
  \country{USA}
}
\email{shaz@fb.com}          

\thanks{*Author names are in alphabetical order}
\begin{abstract}
The Move language provides abstractions for programming with digital assets via a mix of value semantics and reference semantics. Ensuring memory safety in programs with references that access a shared, \emph{mutable} global ledger is difficult, yet essential for the use-cases targeted by Move. The language meets this challenge with a novel memory model and a modular, intraprocedural static reference safety analysis that leverages key properties of the memory. The analysis ensures the absence of memory safety violations in all Move programs (including ones that link against untrusted code) by running as part of a load-time \emph{bytecode verification} pass similar to the JVM \cite{jvm} and CLR \cite{clr}. We formalize the static analysis and prove that it enjoys three desirable properties: absence of dangling references, referential transparency for immutable references, and absence of memory leaks.
\end{abstract}



\maketitle

\section{Introduction}


The Move language\cite{move_white} provides abstractions for writing safe
\emph{smart contracts} \cite{szabo_smart_contracts, ethereum}
via a mix of value semantics and reference semantics.
Programmers can use value semantics and the \code{move} operator
to transfer ownership of an asset to another user or to a procedure.
For example, a procedure signature like \code{fun buy(c: Coin): Car}
intuitively says: ``if you give me ownership of a coin,
I will give you ownership of a car''.
By contrast, references enable programmers to temporarily share a
value for reading or writing.
The signature \code{fun register(c: &Car, fee: Coin): Registration} says
``if you show me that you own a car and pay a fee, I will give you a registration for the car''.

However, careless mixing of value and reference semantics can lead to memory safety issues.
Code like \code{let x = 5; return &x} creates a dangling reference to local memory (as in C).
In addition, the \code{move} operator introduces new kinds of reference errors:
\code{let y = move x; *x} (``use-after-move'', similar to a null dereference)
and \code{let y = &x; let z = move x;} (\code{y} is a dangling reference to moved memory).

Smart contract languages like Move must prevent memory safety issues \emph{by design}
because they need to support deterministic execution in an adversarial environment.

\paragraph{Deterministic execution}
A blockchain is a replicated state machine \cite{Lamport:1984:UTI:2993.2994, Schneider:1990:IFS:98163.98167}
where replicas are known as \emph{validators}.
Users in the system send transactions (i.e., programs in a language like Move) to a validator that advances the state machine.
The validators execute a consensus protocol (e.g., \cite{nakamoto, libra_consensus_white})
to agree on the ordering and results of executing transactions.
State-machine replication requires determinism;
if execution is nondeterministic, validators may not achieve consensus and the system cannot make progress.
Since violation of memory safety typically manifests as undefined behavior,
the resulting nondeterminism could stall the entire system.

\paragraph{Adversarial environment}
Smart contracts deployed on a blockchain store digital assets with real-world value,
but must tolerate arbitrary interactions with untrusted code running in the same address space.
Even if the deployed code is completely safe on its own,
an attacker can intentionally write memory-unsafe code that attempts to break the integrity of the deployed code.
Thus, Move must prevent memory safety issues in \emph{both} the deployed code and the attacker code.

\paragraph{Existing approaches are inadequate}
Unfortunately, traditional approaches for ensuring memory safety are not suitable for Move's deployment model.
For example, Move cannot rely on a source language with a strong type system (e.g., Rust, OCaml)
to prevent memory safety issues because Move \emph{bytecode} is stored and executed on the ledger.
An attacker that wishes to subvert the source-level type system can write and deploy bytecode directly.

Similarly, Move cannot utilize GC for safe memory management.
Like other blockchain languages, Move uses \emph{gas metering}\cite{DBLP:journals/cacm/GrechKJBSS20} to
provide a deterministic metric for execution cost of contracts.
The gas cost of a program must closely track its execution time to avoid
denial-of-service vulnerabilities\cite{DBLP:conf/ndss/0002L20}.
The unpredictable nature of GC does not mix well with the precise accounting
required by gas metering.

\paragraph{Contributions}
Move must support a rich programming model.
An account-based blockchain is a shared, \emph{mutable} global ledger.
A single transaction can mutate an arbitrary number of accounts in the ledger
(e.g., by sending funds to each).
Move must expose a programming model that supports mutable references to global state,
but without sacrificing reference safety.

Move addresses safe programming of smart contracts by chaining together three new ideas.
First, memory management in Move is built atop a forest of tree-shaped values.
The roots of this forest are either local variables on the call stack or global memory
indexed by a statically-known type and dynamically determined account address.
Move allows the creation of references to values and embedded sub-values,
but does not allow references to be embedded inside a value.
This design ensures that a reference can be canonically represented as a rooted path of fields.

Second, Move imposes the abstraction of a directed and acyclic borrow graph over
the forest of values comprising the program memory.
The nodes in this borrow graph are abstract locations representing values
on the call stack, values indexed by a particular type in global memory,
or references stored on the call stack or operand stack.
Each edge in the borrow graph is labeled by a relative path indicating
an ownership transfer along that path.

Third, Move provides an intraprocedural static analysis to
automatically check that a Move bytecode program adheres
to the ownership discipline enforced by the borrow graph abstraction.
This analysis ensures important properties
---no memory leaks, no dangling references, and referential transparency---
on Move bytecode programs.

We have implemented our analysis in the Move bytecode verifier
which is run whenever new bytecode is loaded into the
\Libra blockchain\cite{libra_blockchain_white},
in a manner similar to the JVM~\cite{jvm} and CLR~\cite{clr}.
Therefore, runtime memory-safety properties are enforced on all loaded bytecode
regardless of provenance.
Our implementation is fast; we report numbers in $\S \ref{sec:evaluation}$.

We have also implemented our analysis in the compiler for the Move source language.
The Move language is being used by \Libra developers to implement
the rules governing the \Libra Payment Network.
Anecdotal experience indicates that the source borrow checker
is a useful tool that helps developers write safe code.

\section{Language Design for Reference Safety}
\label{sec:language_design}

In this section, we will first give an overview of memory management in the Move language with emphasis on the key feature for facilitating static analysis: tree-shaped values that enable us to represent reference values as structured paths (\S \ref{sec:storage_and_refs}).

Next, we present a static analysis that builds on these features to prevent dangling references. We organize our discussion around the three primary challenges our analysis faces: preventing dangling references to memory in the same procedure (\S \ref{sec:dangling_local_refs}), to memory in different procedures (\S \ref{sec:procedure_borrows}), and to global memory (\S \ref{sec:acquires_analysis}). The global reference analysis leverages Move's type-indexed global memory and encapsulation features to enforce a global property with intraprocedural checks.

Each of the analysis sections contain examples of bad Move code that must be rejected. The analysis operates on Move bytecode, but we will write our examples in Move source code for readability.
The table below summarizes how source code instructions compile to stack-based bytecode instructions.
\[
  \begin{tabular}{|ll|}\hline
    \code{x = &y} & $\ao{\borrowlocCmd}{y}; \ao{\storelocCmd}{x}$ \\
    \code{x = &y.f} & $\ao{\borrowfieldCmd}{f,y}; \ao{\storelocCmd}{x}$ \\
    \code{x = *y} & $\ao{\copylocCmd}{y}; \readrefCmd; \ao{\storelocCmd}{x}$ \\
    \code{*x = y} & $\ao{\movelocCmd}{y}; \ao{\borrowlocCmd}{x}; \writerefCmd$ \\
    \hline
  \end{tabular}
\]
\noindent
Although the source language abstracts away some bytecode features such as the operand stack, we have chosen examples that capture the essence of reference issues in the bytecode.

\subsection{Memory management}
\label{sec:storage_and_refs}

Move has three kinds of storage:
\begin{enumerate}
\item \textbf{Procedure-local variables} Each procedure frame in the call stack has a fixed set of local variables that are uninitialized at the beginning of a procedure. Initialized variables can store values and references.
\item \textbf{Shared operand stack} All procedures in the call stack share a single operand stack that can store both values and references. Procedures can use the operand stack both for local computation and to share arguments/return values with other procedures. At the beginning of program execution, the call stack holds one frame and the operand stack is empty. The same conditions must hold for a program to terminate without an error.
\item \textbf{Global storage} Move has no global variables, no heap, and cannot access the filesystem or network. However, programs can access persistent data by reading from and writing to shared global storage that persists across program executions. Global storage is organized as a partial map from 16 byte \emph{account addresses} to record values: ($\textsf{Addr} \times \textsf{Type}) \rightharpoonup \textsf{Record}$. \S~\ref{sec:acquires_analysis} explains the design of and motivation for the global storage in more detail.
\end{enumerate}





\paragraph{Values are Tree-Shaped}
Move has primitive data values, nominally typed records, resizable vectors, and reference values. Move programmers can create references that point into local variables or into global storage, but not into the operand stack. Reference types are tagged with \emph{permissions}: either mutable (written \code{&mut T}) or immutable (written \code{&T}). References to other references (e.g., \code{&&u64}) are not allowed.

Both records and vectors can store primitive values, and other records/vectors, but not references. Global storage can hold records, but not references.
This ensures that non-reference values and the global storage are always tree-shaped.


\paragraph{References are Structured Paths}
In a byte-addressable memory, a reference value is a dynamically determined index into the array of memory. This unstructured representation makes it difficult to reason about the relationship between two different reference values--e.g., ``can writing through reference \code{i} change the memory pointed to by reference \code{j}?''

By contrast, Move storage, values, and reference-related instructions are designed to ensure that a reference can be represented as a structured \emph{access path}~\cite{Jones-al:POPL79}. For example: say we have a record value \code{\{ f: \{ g: 1 \}, h: [2,3] \}} of type \code{T}, where the \code{\{} syntax denotes a record and the \code{[} syntax denotes a vector. If this value is stored in a local variable \code{x}, we can represent a reference to the value stored by field \code{g} as the path \code{x/f/g}. Similarly, we can represent a reference stored at index 1 of the vector in field \code{h} as \code{x/h/1}. If the same record value is in global storage at account address \code{A}, we can represent these paths as \code{A/T/f/g} and \code{A/T/h/1}.

A path is a canonical representation for a particular location in memory. Syntactically distinct paths refer to distinct memory locations. In addition, the structured nature of paths introduces a partial order on reference values. Two reference values either have a prefix relationship (e.g., \code{x/f} is a prefix of \code{x/f/g}), an extension relationship (e.g., \code{x/f/g} is an extension of \code{x/f}), or are incomparable (e.g., \code{y/f/g} and \code{x/f/g}). Writing to a path cannot change the memory named by incomparable paths. As we will see, our reference safety analysis takes advantage of these nice properties to simplify static reasoning about code that uses references.

\subsection{Preventing Dangling References to Locals}
\label{sec:dangling_local_refs}

We begin our discussion of the reference safety analysis with a pair of code snippets that exemplify the problems the analysis must prevent. The code comments in the snippet show the abstract state of the analysis, but we will ignore them at first.

The program below creates a reference \code{r} to the \code{f} field of formal parameter \code{c}, \emph{moves} the value stored in \code{c} into \code{x} and then reads \code{r}. The \code{move} assignment works by assigning the value bound to \code{c} to \code{x} and then ``clearing'' \code{c} by assigning it to $\bot$.
\begin{lstlisting}[basicstyle=\scriptsize\ttfamily,language=Move]
fun dangle_after_move(c: Coin) {
  let r = &c.f; // $\borrow{\texttt{c}}{\texttt{f}}{\texttt{r}}$
  let x = move c; // $\reject$
  let y = *r; // read from dangling ref!
}
\end{lstlisting}

The ability (and in some cases, requirement) to move values instead of copying them is a key feature of the Move language---it prevents ``double spending'' of monetary values like \code{c}. However, the \code{move} causes the reference value stored in \code{r} to become dangling.

Similarly, if the programmer creates a reference to a value that is overwritten via a destructive update, a dangling reference may result. The snippet below creates a vector \code{v} of size 1, acquires a reference \code{ref} into index 0 of the vector, and then reassigns \code{v} to an empty vector via the write to \code{v_ref}. As a result, the write \code{*ref} accesses a dangling reference.
\begin{lstlisting}[basicstyle=\scriptsize\ttfamily,language=Move]
let v = Vector::singleton(5);
let v_ref = &mut v; // $\borrow{\texttt{v}}{\varepsilon}{\texttt{v\_ref}}$
let ref = &mut v[0]; // $\borrow{\texttt{v}}{*}{\texttt{ref}}$, $\borrow{\texttt{v}}{\varepsilon}{\texttt{v\_ref}}$
*v_ref = Vector::empty(); // $\reject$
*ref = 10; // write to dangling ref!
\end{lstlisting}


\paragraph{Ownership and Borrowing}

Our analysis enforces a programming discipline based on ownership to prevent the problems above. A location in memory (either a local variable, stack location, or global key) of type \code{T} is the \emph{owner} of the value it stores. A value of type \code{T} can only be moved or (if \code{T} is a mutable reference) written via its owning location. However, a value stored in a local or global can be \emph{borrowed} from its owner by creating a reference to it (e.g. \code{&v}) or extending an existing reference (e.g. \code{&ref.f}). The analysis will not allow the value to be moved or written until all borrows have ended (i.e., the reference values produced by the borrows have been destroyed). This discipline has a natural recursive structure: if \code{y} borrows from \code{x} and then \code{z} subsequently borrows from \code{y}, \code{x} does not regain ownership until both the \code{y} and the \code{z} borrows have ended.

As we will show in \S \ref{sec:abstractSem}, a program that follows these rules cannot create a dangling reference by writing path $p$ when a reference path that is a strict extension of $p$ exists elsewhere or moving a variable $x$ when a reference path that is an extension of $x$ exists elsewhere. As a side benefit, this discipline ensures referential transparency for immutable references.

\paragraph{Borrow Analysis Mechanics}

The key piece of analysis state is a \emph{borrow graph} where nodes represents references or values and a directed edge $\borrow{A}{p}{B}$ means ``path $A/p$ is borrowed by $B$''. Here, $A/p$ is an abstraction of the path representation described in \S \ref{sec:storage_and_refs} extended with some additional components: $\epsilon$ for a direct borrow of a local variable and $*$ for all suffixes of a path. The analysis adds a borrow edge when a reference value is created and eliminates borrow edges when a reference value is destroyed (e.g., popped off the stack or consumed by an instruction).

Returning to our examples above, each instruction is annotated with the borrow graph after it executes. A $\reject$ indicates that the borrow checker would reject the program after processing the instruction. The first program is straightforward: the \code{&c.f} instruction creates a borrow of \code{c}, then the analysis rejects the program at \code{move c} because there is a borrow edge rooted in \code{c}.

In the second program, the borrow edge $\borrow{\texttt{v}}{\varepsilon}{\texttt{v\_ref}}$ represents that the local \code{v_ref} holds a direct reference to the vector in \code{v}. The next instruction \code{let ref = &mut v[0]} generates the edge $\borrow{\texttt{v}}{*}{\texttt{ref}}$. This is our first encounter with abstraction in the analysis: the $*$ means that the analysis does not know which offsets of \code{v} have been borrowed by \code{ref}. Finally, the analysis chooses to reject the write to \code{v_ref} because it borrows from \code{v} and the \code{*} edge represents an outstanding borrow on \code{v}.

\subsection{Borrow Discipline Across Procedures}
\label{sec:procedure_borrows}

Move procedures can accept reference arguments and return references, which introduces new ways to create dangling references. However, we can extend the borrow discipline and its corresponding analysis to prevent these as well.

First, we consider the problem of returning dangling references. Each procedure below returns a reference value to its caller by pushing the return value on the operand stack (we write $S_i$ for the $i$th operand stack slot) and then executing the \textsf{Ret} instruction (not shown). The first two procedures in the snippet below return dangling references to local variables of the procedure, but the second two return safe references to memory that will outlive the procedure.
\begin{lstlisting}[basicstyle=\scriptsize\ttfamily,language=Move]
fun ret_local(): &u64 { let x = 7; &x /* $\borrow{\texttt{x}}{\varepsilon}{S_0}$ */ } $\reject$
fun ret_param(x: u64): &u64 { &x /* $\borrow{\texttt{x}}{\varepsilon}{S_0}$ */ } $\reject$
fun ret_ref_param(x: &u64): &u64 { x /* $\emptygraph$ */ }
fun ret_borrowed_param(s: &S): &u64 { &s.f /*$\borrow{\texttt{s}}{\texttt{f}}{S_0}$*/ }
\end{lstlisting}

Each procedure shows the borrow graph before the procedure returns.
The analysis will not allow a return value to have any borrow of a local
variable (\code{ret_local}) or a formal with a non-reference type (\code{ret_param}).
The \code{ret_ref_param} procedure is safe because it has no borrows---it returns a copy of
a reference parameter. Finally, \code{ret_borrowed_param} has a borrow of formal \code{s},
but this is ok because \code{s} is a reference parameter rather than a local.
The analysis has access to a procedure's type information, and it also tracks the type of
each stack location in a separate abstract domain.

\paragraph{Procedure Calls Require Ownership}
Lastly, we consider the problem of handling procedure calls. Our static analysis is modular, so it must soundly summarize the effects of a procedure call with no information other than the callee's type signature. A reference value returned by a callee is conservatively assumed to be borrowed from all of the procedure's reference arguments, with extra care to avoid conflating immutable and mutable references (see the \textsf{Call} rule in \S \ref{sec:abstractSem} for details).

A subtle consequence of the borrow discipline described in the previous section is that every non-reference value pushed on the operand stack has no outstanding borrows. Since procedures pass arguments to their callees on the operand stack, this ensures that a caller cannot retain references to a value passed to a callee. With this possibility out of the way, the only danger that remains is a dangling reference caused by a reference value written in a callee. To prevent this, the analysis enforces a single rule: an argument with a mutable reference type must not have any outstanding borrows.

The two examples below show unsafe call sites. In both cases, the arguments are evaluated left to right and pushed on the operand stack before the \textsf{Call} instruction. The borrow graphs are at the program point before this instruction. The analysis rejects each example because stack slot $S_1$ is passed as an argument, but is borrowed by $S_0$.
\begin{lstlisting}[basicstyle=\scriptsize\ttfamily,language=Move]
a(&mut x, &mut x) // $\borrow{\texttt{x}}{\varepsilon}{S_1}, \borrow{S_1}{\varepsilon}{S_0}$ $\reject$
b(&mut y, &mut y.f) // $\borrow{\texttt{y}}{\varepsilon}{S_1}, \borrow{S_1}{\texttt{f}}{S_0}$ $\reject$
\end{lstlisting}

In addition to preventing dangling references, this strategy for handling calls ensures a very useful property: \textbf{a mutable reference parameter cannot alias with any other parameter}! Eliminating aliased mutable data allows Move programmers to write procedures without defensive checks for aliasing and greatly facilitates precise and scalable static program verification in the Move Prover \cite{moveprover}.

\subsection{Dangling References to Global Memory}
\label{sec:acquires_analysis}

We conclude our informal presentation of the analysis by considering the thorny problem of allowing mutable access to global memory while preventing dangling references. At first blush, this might seem impossible to do with a modular analysis--the whole point of global memory is that you can access it from anywhere! However, Move's global memory instructions are carefully designed to enable local reasoning about the safety of global memory accesses. The two analysis extensions described in this section build on this design to prevent dangling references to global memory using the existing borrow analysis machinery.

\paragraph{Global Memory Indexed By Encapsulated Types}
Move's global memory is a partial map indexed by a pair of a statically chosen record type \code{T} and a dynamically chosen account address \code{a}. A value stored at key \code{(T, a)} is a record of type \code{T}. Record types are declared in \emph{modules} consisting of type and procedure declarations.
The following table summarizes the global state operations available in Move.
\[
  \begin{tabular}{|ll|}\hline
    \code{move_to<T>(a, T)} & Publish \code{T} at address \code{a} \\
    \code{move_from<T>(a): T} & Remove \code{T} from \code{a} \\
    \code{borrow_global<T>(a): \&mut T} & Get mutable ref to \code{T} at \code{a} \\
    \hline
  \end{tabular}
\]
Programmers can publish a value of type \code{T} to an address in global state, remove the value of type \code{T} stored at an address, and acquire references to a value already published in global state. Modules encapsulate access to their declared types; in particular, the global state access operations can only be used on a type declared \emph{inside the current module}.

The decision to include a type in these operations simplifies static reasoning about aliasing of locations in global memory. Two global access operations involving keys \code{(T1, a1)} and \code{(T2, a2)} can only touch the same memory if \code{T1} = \code{T2} and \code{a1} = \code{a2}. Combined with type encapsulation, this means that global accesses in distinct modules touch distinct memory by construction. Without this property, a local reference safety analysis would not be practical.

\paragraph{Abstracting Dynamic Global Accesses}
Each instruction that accesses global memory indexes into it using an \code{address} value chosen at runtime. This means that code like:
\begin{lstlisting}[basicstyle=\scriptsize\ttfamily,language=Move]
fun address_aliasing(a1: address, a2: address) acquires T {
  let t_ref = borrow_global<T>(a1); // $\borrow{\texttt{T}}{\varepsilon}{\texttt{t\_ref}}$
  let t = move_from<T>(a2); // $\reject$
  *t_ref = ... // accessing a dangling ref
}
\end{lstlisting}
may create a dangling reference if \code{a1} and \code{a2} are bound to the same address value. Similarly, performing a \code{move_to<T>(a, T} while a reference created by a \code{borrow_global<T>(a)} is still active creates a dangling reference.

Rather than attempting to reason about address equality, our analysis conservatively assumes that all global accesses indexed by type \code{T} touch the same address.
This decision suggests a simple extension to the borrow graph: a \code{T} node that abstracts all concrete cells in global memory keyed by type \code{T}. In the example above, the analysis introduces a borrow from \code{T} at the \code{borrow_global} instruction and rejects the program at \code{move_from<T>} because there is an active borrow on \code{T}.
This is exactly how the analysis deals with programs like \code{let x = &y; let z = move y} only with \code{T} in place of \code{x} and global access instructions in the place of local ones.

\paragraph{Global Accesses Across Procedure Boundaries}
A similar, but more insidious problem arises if the global accesses occur in procedures with a caller/callee relationship:
\begin{lstlisting}[basicstyle=\scriptsize\ttfamily,language=Move]
fun borrow_then_remove_bad(a: address) acquires T {
  let t_ref = = borrow_global<T>(a); // $\borrow{\texttt{T}}{\varepsilon}{\texttt{t\_ref}}$
  let t = remove_t(a); $\reject$
  *t_ref = ...// accessing a dangling reference
}
fun remove_t(a: address): T acquires  T {
  move_from<T>(a)
}
\end{lstlisting}

However, the \code{acquires} annotation on \code{remove_t} allows the borrow analysis to reject this program--an invocation of a procedure marked with \code{acquires T} is treated as a move of \code{T}.

A separate intraprocedural analysis checks that any procedure containing a \code{borrow_global<T>} or \code{move_from<T>} instruction is annotated with \code{acquires T}. Any procedure that invokes an \code{acquires T}-annotated procedure in the same module must also be annotated. A procedure that calls an \code{acquires} procedure declared in a different module need not be annotated.

\paragraph{Global References Cannot Be Returned}

The scheme described so far prevents dangling references to a global memory cell of type \code{T} inside the module that declares \code{T}. Only one issue remains: leaking a reference to a global cell of type \code{T} outside of the module where \code{T} is declared:
\begin{lstlisting}[basicstyle=\scriptsize\ttfamily,language=Move]
module M1 {
fun f(a: address): &mut T acquires T { borrow_global<T>(a) }
fun g(a: address): T acquires T { move_from<T>(a) }
}
module M2 {
fun bad(a: address) {
  let ref = M1::f(a);
  let t = M1::g(a); // ref now dangling!
}}
\end{lstlisting}
The possibility of this leakage undermines our efforts to modularize reasoning about reference invalidation. Thus, our analysis prevents it by banning returns of references to global memory. This happens implicitly by following the return discipline described in \S \ref{sec:procedure_borrows}: only borrows of reference parameters may remain on the stack when a procedure returns, and a global is not a reference parameter.

\section{\move Operational Semantics}
\label{sec:concrete}

\begin{figure}
\begin{centerframe}
\pcrule[]
  {
    z = (\proc, \pc, L)\sep
    \progloc{\proc}{\pc}=\ao{\callCmd}{\proc'}
  }
  {\program}{\st{\stackht{z}{P},\stackht{\overline{v}}{S},M}}{\st{\stackht{(\proc', 0, \reverse{\overline{v}})}{\stackht{z}{P}},S,M}}\\
\vspace*{1mm}
  \pcrule[Emp]
  {
    z = (\proc, \pc, L)\sep
	  \progloc{\proc}{\pc}=\returnCmd\sep
    M' = M - img(L)
  }
	{\program}{\st{z,[],M}}{\st{[],[],M'}}\\
\vspace*{1mm}
\pcrule[Ret]
{
  z = (\proc, \pc, L)\sep
  \progloc{\proc}{\pc}=\returnCmd\sep
  M' = M - img(L)
}
{\program}{\st{\stackht{z}{\stackht{(\proc', \pc', L')}{P}},S,M}}{\st{\stackht{(\proc', \pc'+1, L')}{P},S,M'}}\\
\vspace*{1mm}
\pcrule[]
  {
	  \progloc{\proc}{\pc}=\ao{\branchCmd}{\pc_1, \pc_2}\sep
    \pc' = \ite{v}{\pc_1}{\pc_2}
  }
  {\program}{\st{\stackht{(\proc, \pc, L)}{P},\stackht{v}{S},M}}{\st{\stackht{(\proc, \pc', L)}{P},S,M}}\\
\vspace*{2mm}
  \pcrule[\cmdname{Execute}]
  {
    \pcstep{\progloc{\proc}{\pc}}{\st{L,S,M}}{\st{L',S',M'}}
  }
	{\program}{\st{\stackht{(\proc, \pc, L)}{P},S,M}}{\st{\stackht{(\proc, \pc + 1, L')}{P},S',M'}}\\
\vspace*{1mm}
\pcrule[]
  {
    v = \ite{\lread{L}{x}\in\dom{M}}{\mread{M}{\lread{L}{x}}}{\lread{L}{x}}
  }
  {\ao{\movelocCmd}{x}}
  {\st{L,S,M}}
  {\st{\mdel{L}{x},\stackht{v}{S},\mdel{M}{\lread{L}{x}}}}\\
\vspace*{1mm}
\pcrule[]
  {
    v = \ite{\lread{L}{x}\in\dom{M}}{\mread{M}{\lread{L}{x}}}{\lread{L}{x}}
  }
  {\ao{\copylocCmd}{x}}
  {\st{L,S,M}}
  {\st{L,\stackht{v}{S},M}}\\
\vspace*{1mm}
\pcrule[Ref]
  {
  }
  {\ao{\storelocCmd}{x}}
  {\st{L,\stackht{r}{S},M}}
  {\st{\lset{L}{x}{r},S,M}}\\
\vspace*{1mm}
\pcrule[Val]
  {
    \lread{L}{x} = c \vee  (x\not\in\dom{L} \wedge c\notin\dom{M})
  }
  {\ao{\storelocCmd}{x}}
  {\st{L,\stackht{v}{S},M}}
  {\st{\lset{L}{x}{c},S,\mset{M}{c}{v}}}\\
\vspace*{1mm}
\pcrule[]
  {
  }
  {\ao{\borrowlocCmd}{x}}
  {\st{L,S,M}}
  {\st{L,\stackht{\rft{\lread{L}{x}}{[]}}{S},M}}\\
\vspace*{1mm}
\pcrule[]
  {
    \lread{L}{x} = \rft{c}{p}
  }
  {\ao{\borrowfieldCmd}{f,x}}
  {\st{L,S,M}}
  {\st{L,\stackht{\rft{c}{p::f}}{S},M}}\\
\vspace*{1mm}
  \pcrule[]
  {
  }
  {\freezerefCmd}
  {\st{L,\stackht{r}{S},M}}
  {\st{L,\stackht{r}{S},M}}\\
\vspace*{1mm}
\pcrule[]
  {
  }
  {\readrefCmd}
  {\st{L,\stackht{\rft{c}{p}}{S},M}}
  {\st{L,\stackht{\mread{M}{c}[p]}{S},M}}\\
\vspace*{1mm}
\pcrule[]
  {
    r = \rft{c}{p}\sep
    v=\mread{M}{c}
  }
  {\writerefCmd}
  {\st{L,\stackht{v'}{\stackht{r}{S}},M}}
  {\st{L,S,\mset{M}{c}{v[p := v']}}}\\
\vspace*{1mm}
\pcrule[]
  {
    t = \record{(f_{i},\_)\mid 1 \leq i \leq n}\sep
    v = \record{(f_{i},v_{i})\mid 1 \leq i \leq n}
  }
  {\ao{\packCmd}{t}}
  {\st{L,\stackht{v_{1}\cons\cdots\cons v_{n}}{S},M}}
  {\st{L,\stackht{v}{S},M}}\\
\vspace*{1mm}
\pcrule[]
  { v = \record{(f_{i},v_{i})\mid 1 \leq i \leq n}
  }
  {\unpackCmd}
  {\st{L,\stackht{v}{S},M}}
  {\st{L,\stackht{v_{1}\cons\cdots\cons v_{n}}{S},M}}\\
\vspace*{1mm}
\pcrule[]
  {
  }
  {\popCmd}
  {\st{L,\stackht{\_}{S},M}}
  {\st{L,S,M}}\\
\vspace*{1mm}
\pcrule[]
  {
  }
  {\stackopCmd}
  {\st{L,\stackht{v_{1} \cons \cdots \cons v_{n}}{S},M}}
  {\st{L,\stackht{{\stackop(v_1,\dots,v_n)}}{S},M}}
\caption{\label{fig:semantics}Operational semantics}
\end{centerframe}
\end{figure}

In this section, we formalize the operational semantics of a subset of the \move language chosen
to illustrate the key challenges of reference safety.

\paragraph{Partial functions and lists}
We use partial functions to represent record values and for mappings that are parts of semantic states.
Lists are used for sequences of field accesses and for the stack component of semantic states.

Following common convention,
if $f:A\to B$ is a partial function, then $\dom{f}$ is its domain and $\img{f}$ is its range.
We write $\lset{f}{a}{b}$ for the function that is the same as $f$ on every argument except $a$,
and which maps $a$ to $b$.
Similarly, $\mdel{f}{a}$ is the partial function equivalent to $f$ except that it is undefined at $a$.

We write $[]$ for the empty list and $e::l$ for the result of placing $e$ at the front of list $l$.
Similarly, $l::e$ is the list with $e$ appended to $l$ and, by slight overloading of notation,
$l::l'$ is the concatenation of lists $l$ and $l'$.

\paragraph{Types}
Let $\primtype$ be the set of primitive types, including
\booleantype (of Boolean values), \integertype (of integers),
and \addresstype (of account addresses).
Let $\fn$ be a fixed, finite set of \emph{field names}.
The set $\valtype$ of value types is the least set satisfying:
(1)~$\primtype \subseteq \valtype$; and
(2)~if $f_1\dots f_n\in\fn$ are pair-wise distinct and $t_1\dots t_n\in \valtype$, then
  $\{(f_i,t_i)|1\le i \le n\} \in \valtype$.
Let $\immreftype = \{ \imm{t} \mid t \in \valtype \}$,
$\mutreftype = \{ \mut{t} \mid t \in \valtype \}$,
$\reftype = \immreftype \cup \mutreftype$, and
$\type = \valtype \cup \reftype$.

\paragraph{Values}
We define the set of values used in computation inductively.
Let $\primval$ be the set of primitive data values.
The set $\val$ of values is the least set satisfying:
(1)~$\primval \subseteq \val$;
(2)~if $f_1\dots f_n\in\fn$ are pair-wise distinct and $v_1\dots v_n\in \val$, then
$\{(f_i,v_i) | 1\le i \le n\} \in \val$.
The judgment $\typeof{v}{t}$, indicating that value $v$ has type $t$, is defined in the natural way.

\paragraph{Paths}
A path is a possibly empty list of field names,
representing a sequence of field selections.
A path represents a way to start at any location in memory and
follow a sequence of field selections to reach a value.
It is helpful to visualize a value as a labeled tree whose
edges are labeled by field names.
A primitive value is a tree consisting of a single leaf.
The tree associated with a record value consists of a node
and a subtree for each record component.
If $r$ is a record value, then for each $(f,v)\in r$,
the edge from $r$ to the subtree for $v$ is labeled by $f$.

Two useful operations on values are
(1)~the subtree $v[p]$ of $v$ located at path $p$, and
(2)~the tree $v[p:=v']$ obtained by replacing the subtree
at path $p$ with tree $v'$.
Using the association between values and trees, $v[p]$ is the
subtree reached by path $p$ from root $v$,
and $v[p:=v']$ is the tree with the subtree $v[p]$ replaced with $v'$.

\paragraph{Concrete states}
The set $\cl$ is an uninterpreted set of locations;
the memory of a program is a partial map from $\cl$ to $\val$.
A {\em reference} $r$ is a pair $\rft{c}{p}$, where $c \in \cl$ and $p$ is a path.
A \emph{concrete state} $\concState$ is a tuple $\st{P,S,M}$ where:
(1)~\textit{Call stack} $P$ is a list of frames.
Each frame is a triple comprising a procedure $\proc$,
a program counter $\pc$, and a local store $L$ that maps variables to locations or references.
(2)~\textit{Operand stack} $S$ is a list of values and references.
(3)~\textit{Memory} $M$ maps locations to values.

\newcommand{\byteop}{\mathit{op}}

A program \program maps each procedure $\proc$ to a tuple $\program[\proc]$ which contains a nonempty sequence of bytecodes
 $\bcodes{\proc}$, and a sequence of input/output types $\ins{\proc}$ and $\outs{\proc}$.

\paragraph{Semantics}
Figure~\ref{fig:semantics} shows the formal operational semantics.
The first four rules capture the execution of a program $\program$ as a transition
relation $\program \vdash \st{P,S,M} \rightarrow \st{P',S',M'}$.
Each of these rules looks up the instruction $\bcodes{\proc}[\pc]$
pointed to by the top frame on the call stack, executes the instruction,
and updates the state accordingly.
\callCmd establishes a new frame at the top of the call stack.
\returnCmd tears down the frame at the top of the call stack and increments the
program counter for the new topmost frame.
\ao{\branchCmd}{\pc_1,\pc_2} consumes a Boolean value from the top of the operand stack
and updates the program counter to $\pc_1$ if the value is true and to $\pc_2$ otherwise.
The \cmdname{Execute} rule is a wrapper for executing the remaining instructions.

The \cmdname{MoveLoc}, \cmdname{CopyLoc}, and \cmdname{StoreLoc} instructions move or copy values between local variables
and the operand stack;
\cmdname{BorrowLoc}, \cmdname{BorrowField}, \cmdname{ReadRef},
and \cmdname{WriteRef} to operate on reference values stored on the operand stack.
The \cmdname{FreezeRef} rule converts a mutable reference into an immutable one.
This instruction is a no-op in the concrete semantics, but performs several important checks in the analysis (see \Cref{sec:abstractSem}).
\cmdname{Pack} and \cmdname{Unpack} to create and destroy record values by binding values on the stack to fields or
pushing values bound to fields onto the stack (respectively).
\cmdname{Pop} destroys a value on the operand stack. This is simple in the concrete semantics, but the analysis must perform
careful bookkeeping when popping references.
The generic \cmdname{Op} instruction represents arithmetic/bitwise operations
that use the operand stack.

The correctness of executions of $\program$ requires that only instructions in the valid
bytecode range are accessed and that each instruction operates on the correct number of
values of appropriate type producing the appropriate result and updating the state without
violating reference safety.
The first requirement is easily handled by a syntactic check that the bytecode sequence
of each procedure is nonempty, branch targets are legal indices,
and the last instruction is either \branchCmd or \returnCmd.
The second requirement is the subject of the next section.

\section{\move Borrow Checker}
\label{sec:abstractSem}
\setlength{\abovedisplayskip}{0pt}
\setlength{\abovedisplayshortskip}{0pt}
\setlength{\belowdisplayskip}{0pt}
\setlength{\belowdisplayshortskip}{0pt}

In this section, we present the guarantees offered by the \move borrow checker on the
subset of the language formalized in \S \ref{sec:concrete}.
Our formalization requires that there is a type annotation on each bytecode of each procedure.
Our analysis will check that the program is well-typed,
i.e., all type annotations are consistent with each other.
The annotations are derived automatically via a simple fixpoint based on abstract interpretation;
this fixpoint is described informally in \S \ref{sec:implementation}.
The annotations on a well-typed program allow us to define
an abstraction function $\mathit{Abs}$ from a concrete state $\st{P,S,M}$ to an abstract state
$\st{\hat{P},\hat{S},B}$ that replaces each concrete value or reference in $P$ ($S$)
with its type in $\hat{P}$ ($\hat{S}$), drops the memory $M$ entirely replacing it
with a borrow graph $B$ that captures the borrow relationships among values and references
in $P$ and references in $S$.
Finally, we show that every concrete state $s$ reachable by program $\program$ is connected to
$\abs{s}$ by an invariant that is sufficient to prove type safety, absence of leaks, and
absence of dangling references.

In the abstract semantics used to formalize static checking, a path may optionally
end with the distinguished symbol~$*$,
representing an unknown (possibly empty) sequence of additional field selections.
A path is {\em extensible} if it ends in ~$*$ and {\em fixed} otherwise.
We include both kinds of paths in the definitions of path operations below.
We write $p \leq q$ if path $p$ is a prefix of path $q$.
We write $\concat{p}{q}$ for \emph{path concatenation}
which is ordinary concatenation of $p$ and $q$ if $p$ is fixed and
$p$ otherwise.

\paragraph{Borrow graph}
An {\em abstract location} is either $\psloc(x,y)$ for some non-negative integers $x$ and $y$
or $\osloc(x)$ for some non-negative integer $x$.
Let $\absL$ be the set of all abstract locations.
The abstract location $\psloc(x,y)$ represents the contents of the local variable
$y$ in the call stack frame at position $x$.
The abstract location $\osloc(x)$ represents the contents of the operand stack at position $x$.
In both cases, we count up from the bottom of the stack.
An abstract location can be looked up in a concrete state to return either a reference
(both call stack and operand stack), a value (only operand stack),
or a concrete location in $\cl$ (only call stack).

For abstract locations $x,y \in \absL$ and path $p$,
the assertion $\Borrow(x,p,y)$, called a borrow edge, is interpreted in a concrete state $\concState$
to indicate that $x.p$ \textit{is borrowed by} $y$.
If $p$ is fixed, the reference in $\concState$ at position $y$ has jurisdiction
over $m.p$, where $m$ is the reference or concrete location at position $x$
in $\concState$.
If $p$ is extensible, then $x.q$ is borrowed by $y$ for some extension $q$ of $p$.
In our static analysis, we therefore assume (to preserve soundness) that $y$ has
jurisdiction over each $x.q$, for any extension $q$ of the path $p$.

A \emph{borrow graph} $B$ is a collection of borrow edges.
An edge $\Borrow(x, p, y)$ is {\em subsumed} by edge $\Borrow(x,q,y)$ if
either $p=q$ or $q = r*$ and $r \leq p$.
If $G$ and $H$ are borrow graphs, we write $G \below H$,
if every edge $\Borrow(x,p,y)$ in $G$ is subsumed by some edge $\Borrow(x,q,y)$ in $H$.
Note that this definition allows additional edges in $H$ that are not implied by edges in $G$.
Semantically, if $G \below H$, then $G$ imposes every restriction on concrete execution that is
expressed by $H$, and possibly additional restrictions.
As a result, we will see that every state that satisfies $G$ also satisfies $H$.

\paragraph{Abstract state}
An \emph{abstract state} $\absState$ is a tuple $\st{\hat{P},\hat{S},B}$
with three components.
The first component $\hat{P}$, an \textit{abstract call stack} that matches
the concrete call stack,
is a list of triples, each comprising a procedure name \proc,
a program counter $\pc$, and a partial map $\hat{L}$ from variables to $\type$.
The stack $\hat{P}$ defines a set of stack positions
$\emph{Pos}(\hat{P})= \{\psloc(x, y) \mid 0 \leq x < \len{\hat{P}} \wedge y \in \hat{P}[x].\hat{L}\}$,
with $\psloc(x, y)$ indicating the $y$-th local variable in the $x$-th call frame of the stack.
The second component $\hat{S}$, an \textit{abstract operand stack} that matches
the concrete operand stack, is a list of types.
The stack $\hat{S}$ defines a set of meaningful stack positions
$\emph{Pos}(\hat{S})=\{\osloc(x) \mid 0 \leq x < \len{\hat{S}}\}$.
Treating the abstract state $\absState$ as a partial map from positions
$\dom{\absState} = \emph{Pos}(\hat{P}) \cup \emph{Pos}(\hat{S})$ to types,
we let $\absState(\psloc(x, y))$ be the type $\hat{P}[x].\hat{L}[y]$
and $\absState(\osloc(x))$ be the type $\hat{S}[x]$.
The third component $B$ of the abstract state is a borrow graph with edges connecting nodes from
$\dom{\absState}$.

\paragraph{Local abstract state}
Since abstract states represent type information, abstract execution of an imperative program
instruction is a form of type propagation.
The rules for type propagation of local instructions (\S \ref{sec:propagation})
operate over a local abstract state $\st{\hat{L},\hat{S},B}$.
The first component $\hat{L}$ maps the variables in the local store of the top
frame of the call stack to their types.
The second component $\hat{S}$ contains the types of the values and references for the portion
of the operand stack visible to the execution of the procedure in the top frame.
This component is empty at the starting bytecode at position $0$ and grows and shrinks as
values or references are pushed on and popped off it.
The third component $B$ contains the borrow edges for locations in $\hat{L}$ and $\hat{S}$.

Let $\ls = \st{\hat{L},\hat{S},B}$ be an abstract local state.
We define $\dom{\ls} = \{ \psloc(0,i) \mid i \in \dom{\hat{L}} \} \cup
\{ \osloc(i) \mid i \in 0..\len{\hat{S}} \}$.
The locations in $\dom{\ls}$ are defined using the offset $0$ for the position of its
frame and the bottom of its operand stack.
Later, when we define the abstract state corresponding to a concrete state,
we will show how local abstract states of each frame on the call stack,
can be stitched together by adjusting their locations with respect
to an appropriate offset.

We call $\{ \psloc(0,i) \mid i \in 0..\len{\ins{\proc}} \}$ the input
locations of $\proc$.
The abstract local state $\ls$ is {\em well-formed} for $\proc$ if
(1)~every input location of $\proc$ is in $\dom{\ls}$,
(2)~$\ls.B$ is a directed acyclic graph, and
(3)~for all $\Borrow(x,\_,y) \in \ls.B$, we have $x,y \in \dom{\ls}$, and
$y$ is not an input location of $\proc$.
The propagation on local abstract states (\S \ref{sec:propagation}) uses judgments of the form
$
\proc,\op \vdash \st{\hat{L},\hat{S},B} \rightarrow \st{\hat{L}',\hat{S}',B'}
$
indicating that procedure $\proc$ executing instruction $\op$
from $\st{\hat{L},\hat{S},B}$ results in $\st{\hat{L}',\hat{S}',B'}$.
These rules are designed to ensure that if $\st{\hat{L},\hat{S},B}$ is well-formed for
$\proc$, then $\st{\hat{L}',\hat{S}',B'}$ is also well-formed for $\proc$.

\paragraph{Well-typed programs}
In addition to input types $\ins{\proc}$ and output types $\outs{\proc}$, our analysis requires that each procedure $\proc$
contains an abstract local state $\lstates{\proc}[i]$ for each offset $i \in 0..\len{\bcodes{\proc}}$ that is well-formed for
$\proc$.
Let $\Next{\proc}[i]$ be the (empty, singleton, or doubleton) set of program counters
to which control transfer is possible after executing instruction at position $i$.
\[
\begin{array}{llll}
  \Next{\proc}[i] & = & \{\}, & \ifcmd\ \bcodes{\proc}[i] = \returnCmd \\
                  &   & \{\pc_1, \pc_2\}, & \ifcmd\ \bcodes{\proc}[i] = \ao{\branchCmd}{\pc_1, \pc_2} \\
                  &   & \{\pc+1\}, & \mathit{otherwise}
\end{array}
\]
We write $\st{\hat{L},\hat{S},B} \below \st{\hat{L}',\hat{S}',B'}$ if
$\hat{L} = \hat{L}'$, $\hat{S} = \hat{S}'$, and $B \below B'$.
Thus, every concrete local state represented by $\st{\hat{L},\hat{S},B}$ can also be
represented by $\st{\hat{L'},\hat{S'},B'}$.
A program $\program$ is {\em well-typed}
if for all $\proc \in \dom{\program}$:
\begin{enumerate}
\item $\st{\ins{\proc},[],\{\}} = \lstates{\proc}[0]$.
\item For all $i \in \dom{\bcodes{\proc}}$, there exists $\st{\hat{L},\hat{S},B}$ such that
(a)~$\proc,\bcodes{\proc}[i] \vdash \lstates{\proc}[i] \rightarrow \st{\hat{L},\hat{S},B}$,
and
(b)~$\st{\hat{L},\hat{S},B} \below \lstates{\proc}[j]$ for all $j \in \Next{\proc}[i]$.
\end{enumerate}
The first condition states that initially the operand stack and borrow graph are empty.
The second condition expresses that the restrictions imposed
after executing an instruction continue to the next instruction.

\paragraph{Abstraction function}
While the rules
mention only abstract local states,
they also define abstract execution on (full) abstract states.
We define an abstraction function from a concrete state to an abstract state.
As a concrete execution steps through concrete states, its abstract execution
steps through the corresponding abstract states.

We define the abstract state $\st{\hat{P},\hat{S},B}$ corresponding to a
concrete state $\st{P,S,M}$ of a well-typed program $\program$
by looking only at $P$ and the annotations on $\program$.
The component $\hat{P}$ is obtained by processing each frame
$\st{\proc,\pc,L}$ in $P$ separately,
replacing $L$ with $\lstates{\proc}[\pc].\hat{L}$,
and concatenating the results:
\[
  \begin{array}{l}
    \absframe{P} = \matchcmd\ P \\
    \phantom{xx}\mid [] \rightarrow [] \\
    \phantom{xx}\mid \stackht{\st{\proc,\pc,\_}}{P'} \rightarrow
    \letcmd\ x = \lstates{\proc}[\pc].\hat{L}\ \incmd\ \stackht{\st{\proc,\pc,x}}{\absframe{P'}} \\
    \hat{P} = \absframe{P}
  \end{array}
\]
The definition of $\hat{S}$ is not a straightforward concatenation of the
operand stacks $\lstates{\proc}[i]$ for each position $i \in \dom{P}$
because the operand stack for each frame other than the topmost contains
the arguments for the callee.
These arguments are removed from the operand stack as a result of the call;
consequently, they must be removed prior to concatenation as well.
We first derive $\astates{\proc}[i]$ which removes the callee arguments
from the operand stack if instruction $i$ in $\proc$ is a call.
\[
  \begin{array}[t]{l}
    \ifcmd\ \bcodes{\proc}[i] = \ao{\callCmd}{\proc'} \\
    \hspace{2mm}
    \begin{array}[t]{lll}
      \letcmd \\
      x = \ins{\proc'} \\
      \st{\hat{L},\stackht{\reverse{x}}{\hat{S}},B} = \lstates{\proc}[i] \\
      R = \{(\osloc(\len{\hat{S}}+i), \psloc(1,i)) \mid i \in 0..\len{x}\}\ \incmd \\
      \st{\hat{L},\hat{S},\rename{B}{R}}
    \end{array}\\
    \elsecmd \\
    \hspace{2mm}
    \begin{array}[t]{l} \lstates{\proc}[i] \end{array}
  \end{array}
\]
This derivation uses the operation $\rename{B}{R}$ which renames the nodes of edges
in $B$ according to the bijection $R$.
Here $R$ renames the positions on the operand stack corresponding to callee arguments
to appropriate positions in the next frame with offset $1$.
We now define $\hat{P}$ by concatenating operand stacks obtained by
looking up $\lstates{\proc}$ for the top frame and $\astates{\proc}$ for
all other frames:
\[
  \begin{array}{l}
    \absstack{P} = \matchcmd\ P \\
    \phantom{xx}\mid [] \rightarrow [] \\
    \phantom{xx}\mid \stackht{\st{\proc,\pc,\_}}{P'} \rightarrow
    \letcmd\ x = \astates{\proc}[\pc].\hat{S}\ \incmd\ \stackht{x}{\absstack{P'}} \\
    \hat{S} = \letcmd\ \stackht{\st{\proc,\pc,\_}}{P'} = P,\ x = \lstates{\proc}[\pc].\hat{S}\ \incmd\ \stackht{x}{\absstack{P'}}
  \end{array}
\]
The last component $B$ is defined similarly to $\hat{S}$,
by looking up the borrow graph for each frame,
renaming it appropriately to account for the offset of the frame and its
corresponding operand stack, and taking the union of all such renamed borrow graphs.
The top frame is looked up in $\lstates{\proc}$
but all other frames are looked up in $\astates{\proc}$.
\[
  \begin{array}{l}
    \absgraph{P}{n} = \matchcmd\ P \\
    \phantom{xx}\mid [] \rightarrow [] \\
    \phantom{xx}\mid \stackht{\st{\proc,\pc,\_}}{P'} \rightarrow \letcmd\\
    \phantom{xxxx} B = \ifcmd\ n = \len{P}\ \thencmd\ \lstates{\proc}[\pc].B\ \elsecmd\ \astates{\proc}[\pc].B, \\
    \phantom{xxxx} R = \{ (\psloc(x,y),\psloc(x+\len{P'},y)) \mid \psloc(x,y) \}\ \cup \\
    \phantom{xxxxxxxx}\{ (\osloc(x),\osloc(x + \len{\absstack{P'}})) \mid \osloc(x) \}\ \incmd \\
    \phantom{xxxx}\rename{B}{R} \cup \absgraph{P'}{n} \\
    B = \absgraph{P}{\len{P}}
  \end{array}
\]
The constraints on the annotations of a well-typed program, explained earlier,
allow us to prove that $B$ is acylic,
an important property that we leverage in the proof of reference safety.
Finally, we get $\abs{\st{P,S,M}} = \st{\hat{P},\hat{S},B}$,
where $\hat{P}$, $\hat{S}$, and $B$ are defined as above.

We use the abstraction function to prove critical invariants
about executions of well-typed programs.
These invariants establish useful properties---type agreement, no memory leaks,
no dangling references, and referential transparency.
We state the invariants as a predicate $\inv(\concState,\absState)$
over a concrete state $\concState$ and an abstract state $\absState$.
We use type propagation on local abstract states (\S \ref{sec:propagation})
to prove the following theorem:
\begin{manualtheorem}{1}
\label{type-sound-thm}
    Let program $\program$ be well-typed.
    If $\concState$ is a concrete state with $\inv(\concState,\abs{\concState})$
    and $\program \vdash \concState \rightarrow \concState'$,
    then $\inv(\concState',\abs{\concState'})$.
\end{manualtheorem}
A proof sketch for this theorem is available in the supplemental material.
A corollary is that if $\program$ starts execution in
a concrete state $\concState_0$ such that $\inv(\concState_0,\abs{\concState_0})$ holds,
then $\inv(\concState,\abs{\concState})$ holds for all states
reachable from~$\concState_0$.
Any initial state $\concState_0$ of $\program$ is of the form
$\st{(\proc,0,\overline{x}),[],\{\}}$ representing the beginning of
a transaction that invokes $\proc$ with inputs $\overline{x}$ comprising only values
(no references), empty operand stack, and empty memory.
It is easy to see that $\inv(\concState_0,\abs{\concState_0})$ holds
if $\program$ is well-typed.
We present $\inv$ as the conjunction of four predicates,
$\inva$, $\invb$, $\invc$, $\invd$, described below.

\paragraph{Type Agreement}
Concrete state $\concState = \st{P,S,M}$ and
abstract state $\absState = \st{\hat{P},\hat{S},B}$ are {\em shape-matching} if
(1)~$\len{P} = \len{\hat{P}}$,
(2)~$\len{S} = \len{\hat{S}}$,
(3)~for all $i \in \dom{P}$, we have $P[i].\proc = \hat{P}[i].\proc$,
$P[i].\ell = \hat{P}[i].\ell$, and
$\dom{P[i].L} = \dom{\hat{P}[i].\hat{L}}$.
Intuitively, shape-matching states have the same call stack height, the same  operand stack height,
and agreement between corresponding procedure names, program counters, and set of local variables in each
stack frame.
Shape-matching states $\concState = \st{P,S,M}$  and $\absState = \st{\hat{P},\hat{S},B}$ are
further \emph{type-matching} if for all positions $n$ in the identical sets
of call stack positions $\emph{Pos}(P)=\emph{Pos}(\hat{P})$
or in the identical sets of operand stack positions $\emph{Pos}(S)=\emph{Pos}(\hat{S})$,
we have $\typeof{\concState(n)}{\absState(n)}$.

$\inva(\concState,\absState):$
$\concState$ and $\absState$ are shape-matching and type-matching.

\paragraph{No Memory Leaks}
The following invariant indicates that
(1)~every local variable on the call stack of $\concState$ contains a different location, and
(2)~locations are not leaked, i.e., $\concState.M$ does not contain any location
not present in a local variable.

\paragraph{$\invb(\concState,\absState):$}
The relation
$\{ (n,\concState(n)) \mid n \in \dom{\absState} \wedge \concState(n) \in \cl \}$
is a bijection from its domain to $\dom{\concState.M}$.

\paragraph{No Dangling References}
For shape-matching $\concState = \st{P,S,M}$ and $\absState = \st{\hat{P},\hat{S},B}$,
a borrow edge $\Borrow(m,p,n)$ in $B$
is \emph{realized in} $\concState$ if the path $p$ leads from $\concState(m)$ to $\concState(n)$,
optionally involving additional field selections if $p$ ends in $*$.
More precisely, we say this graph edge is realized if either
\begin{enumerate}
\item $\concState(m) = c$, $\concState(n) = \rft{c}{q'}$, and path $p$ matches $q'$, or
\item $\concState(m) = \rft{c}{q}$, $\concState(n) = \rft{c}{q'}$, and $q.p$ matches $q'$.
\end{enumerate}
Note that the two conditions express the same basic relationship if we identify $c$ and $ \rft{c}{\varepsilon}$.
The following invariant allows us to conclude that every reference is rooted
in a memory location present in some local variable on the call stack.

\paragraph{$\invc(\concState,\absState):$}
\begin{enumerate}
\item $\absState.B$ is acyclic.
\item
For all $n \in \dom{\absState}$ such that $\absState(n) \in \valtype$,
there is no borrow edge in $\absState.B$ coming into $n$.
\item
For all $n \in \dom{\absState}$ such that $\absState(n) \in \reftype$,
there is a borrow edge in $\absState.B$ coming into $n$ that is realized in $\concState$.
\end{enumerate}

\paragraph{Referential Transparency}
We write $\rft{c}{p} \leq \rft{d}{q}$ if $c = d$ and $p \leq q$.
We extend $\leq$ so that $c \leq \rft{c}{p}$ for any~$p$.
The following invariant indicates that the absence of borrow edges out of an
abstract location containing a value or a mutable reference
guarantees that mutation via that abstract location,
either of the stored value or the value pointed to by the mutable reference,
will not invalidate any live reference.

\paragraph{$\invd(\concState,\absState):$}
For any distinct $m,n\in\dom{\absState}$ such that
$\absState(n) \in \reftype$ and $\concState(m) \leq \concState(n)$,
one of the following hold:
\begin{enumerate}
\item
$\absState(m) \in \immreftype$ and $\absState(n) \in \immreftype$.
\item
$\absState(m) \not \in \immreftype$
and there is a path in $\absState.B$ from $m$ to $n$
comprising realized edges in $\concState$.
\item
$\concState(m) = \concState(n)$
and there is a path in $\absState.B$ from $n$ to $m$
comprising realized edges in $\concState$.
\end{enumerate}

\subsection{Propagating Local Abstract States}
\label{sec:propagation}

\begin{figure*}
\begin{centerframe}

\pcrule[]
  {
    x \not \in 0..\len{\ins{\proc}} \sep
    x \in \dom{\hat{L}}\sep
    \lread{\hat{L}}{x} \in \valtype\Rightarrow
    \Borrow(\psloc(0,x),\_,\_) \not \in B\sep
    B' = \rename{B}{\{(\psloc(0,x),\osloc(\len{\hat{S}}))\}}
  }
  {\proc,\ao{\movelocCmd}{x}}
  {\st{\hat{L},\hat{S},B}}
  {\st{\mdel{\hat{L}}{x},\stackht{\lread{\hat{L}}{x}}{\hat{S}},B'}}
  \bigskip

\pcrule[]
  {
    B'= \elim{B}{\osloc(\len{\hat{S}})}
  }
  {\proc,\popCmd}
  {\st{\hat{L},\stackht{\_}{\hat{S}},B}}
  {\st{\hat{L},\hat{S},B'}}
  \hspace{4mm}
\pcrule[Ref]
  {
    x \not \in 0..\len{\ins{\proc}} \sep
    \lread{\hat{L}}{x} \in \reftype\sep
    B' = \rename{\elim{B}{\psloc(0,x)}}{\{(\osloc(\len{\hat{S}}),\psloc(0,x))\}}
  }
  {\proc,\ao{\storelocCmd}{x}}
  {\st{\hat{L},\stackht{t}{\hat{S}},B}}
  {\st{\lset{\hat{L}}{x}{t},\hat{S},B'}}
  \bigskip

\pcrule[Val]
  {
    x \not \in 0..\len{\ins{\proc}} \sep
    x\not\in\dom{\hat{L}} \vee \lread{\hat{L}}{x} \in \valtype \\\\
    \Borrow(\psloc(0,x),\_,\_) \not \in B \sep
    B' = \rename{B}{\{(\osloc(\len{\hat{S}}),\psloc(0,x))\}}
  }
  {\proc,\ao{\storelocCmd}{x}}
  {\st{\hat{L},\stackht{t}{\hat{S}},B}}
  {\st{\lset{\hat{L}}{x}{t},\hat{S},B}}
  \hspace{8mm}
\pcrule[]
  {
    x \in \dom{\hat{L}}\sep
    \lread{\hat{L}}{x} = t \sep t \in \valtype \\\\
    B' = \factor{B}{\psloc(0,x)}{\osloc(\len{\hat{S}})}
  }
  {\proc,\ao{\borrowlocCmd}{x}}
  {\st{\hat{L},\hat{S},B}}
  {\st{\hat{L},\stackht{\mut{t}}{\hat{S}},B'}}
  \bigskip

\pcrule[Mut]
  {
    x \in \dom{\hat{L}}\sep
    \lread{\hat{L}}{x} = \mut{\record{(f,t),\ldots}} \\\\
    B' = \factorf{f}{B}{\psloc(0,x)}{\osloc(\len{\hat{S}})}
  }
  {\proc,\ao{\borrowfieldCmd}{f,x}}
  {\st{\hat{L},\hat{S},B}}
  {\st{\hat{L},\stackht{\mut{t}}{\hat{S}},B'}}
  \hspace{7mm}
\pcrule[Immut]
  {
    x \in \dom{\hat{L}}\sep
    \lread{\hat{L}}{x} = \imm{\record{(f,t),\ldots}} \\\\
    B' = B \cup \{\Borrow(\psloc(0,x),f,\osloc(\len{\hat{S}})\}
  }
  {\proc,\ao{\borrowfieldCmd}{f,x}}
  {\st{\hat{L},\hat{S},B}}
  {\st{\hat{L},\stackht{\imm{t}}{\hat{S}},B'}}
  \bigskip

\pcrule[]
  {
    x \in \dom{\hat{L}}\sep
    B' = \ite{\lread{\hat{L}}{x} \in \reftype}{\factor{B}{\psloc(0,x)}{\osloc(\len{\hat{S}})}}{B}
  }
  {\proc,\ao{\copylocCmd}{x}}
  {\st{\hat{L},\hat{S},B}}
  {\st{\hat{L},\stackht{\lread{\hat{L}}{x}}{\hat{S}},B'}}
  \vspace{2mm}

\pcrule[]
  {
    \Borrow(\osloc(\len{\hat{S}}), \_, \_) \not \in B\sep
    B' = \elim{B}{\osloc(\len{\hat{S}})}
  }
  {\proc,\writerefCmd}
  {\st{\hat{L},\stackht{t}{\stackht{\mut{t}}{\hat{S}}},B}}
  {\st{\hat{L},\hat{S},B'}}
  \bigskip

\pcrule[]
  {
    t = \imm{t'} \vee t = \mut{t'}\\\\
    \forall n.\ \Borrow(\osloc(\len{\hat{S}}), \_, n) \in B \Rightarrow \st{\hat{L},\hat{S}}[n] \in \immreftype
  }
  {\proc,\freezerefCmd}
  {\st{\hat{L},\stackht{t}{\hat{S}},B}}
  {\st{\hat{L},\stackht{\imm{t'}}{\hat{S}},B}}
  \hspace{5mm}
\pcrule[]
  {
    t = \imm{t'} \vee t = \mut{t'}\sep
    B' = \elim{B}{\osloc(\len{\hat{S}})} \\\\
    \forall n.\ \Borrow(\osloc(\len{\hat{S}}), \_, n) \in B \Rightarrow \st{\hat{L},\hat{S}}[n] \in \immreftype
  }
  {\proc,\readrefCmd}
  {\st{\hat{L},\stackht{t}{\hat{S}},B}}
  {\st{\hat{L},\stackht{t'}{\hat{S}},B'}}
  \bigskip

\pcrule[]
  {
    \overline{t} = [t_1,\ldots,t_n]\sep
    t = \record{(f_{i},t_{i})\mid 1 \leq i \leq n}
  }
  {\proc,\ao{\packCmd}{t}}
  {\st{\hat{L},\stackht{\overline{t}}{\hat{S}},B}}
  {\st{\hat{L},\stackht{t}{\hat{S}},B}}
  \hspace{5mm}
\pcrule[]
  {
    \overline{t} = [t_1,\ldots,t_n]\sep
    t = \record{(f_{i},t_{i})\mid 1 \leq i \leq n}
  }
  {\proc,\unpackCmd}
  {\st{\hat{L},\stackht{t}{\hat{S}},B}}
  {\st{\hat{L},\stackht{\overline{t}}{\hat{S}},B}}
  \hspace{5mm}
\pcrule[]
  {
    \typeof{\stackop}{\overline{t} \rightarrow t}
  }
  {\proc,\stackopCmd}
  {\st{\hat{L},\stackht{\overline{t}}{\hat{S}},B}}
  {\st{\hat{L},\stackht{t}{\hat{S}},B}}
  \bigskip

\pcrule[]
  {
    \is = 0..\len{\ins{\proc'}} \sep
    \os = 0..\len{\outs{\proc'}} \\\\
    \forall i \in \is.\ \ins{\proc'}[i] \in \mutreftype \Rightarrow
    \Borrow(\osloc(\len{\hat{S}}+i),\_,\_) \not \in B\\\\
    B_1 = \rename{B}{\{
                      i \in \is \mid
                      (\osloc(\len{\hat{S}}+i), \psloc(1,i))
                  \}}\\\\
    B_2 = \extend{B_1}{\{
                        \psloc(1,i) \mid
                        i \in \is \wedge \ins{\proc'}[i] \in \mutreftype
                    \}}{\{
                        \osloc(\len{\hat{S}}+i) \mid i \in \os \wedge \outs{\proc'}[i] \in \reftype
                    \}}\\\\
    B_3 = \extend{B_2}{\{
                          \psloc(1,i) \mid
                          i \in \is \wedge \ins{\proc'}[i] \in \immreftype
                      \}}{\{
                          \osloc(\len{\hat{S}}+i) \mid
                          i \in \os \wedge \outs{\proc'}[i] \in \immreftype
                      \}}\\\\
    B' = \elim{B_3}{\{ \psloc(1,i) \mid i \in \is \}}
  }
  {\proc,\ao{\callCmd}{\proc'}}
  {\st{\hat{L},\stackht{\reverse{\ins{\proc'}}}{\hat{S}},B}}
  {\st{\hat{L},\stackht{\reverse{\outs{\proc'}}}{\hat{S}},B'}}
  \bigskip

\pcrule[]
  {
    \hat{L} = \ins{\proc} \sep \forall x \in \dom{\hat{L}}.\ \lread{\hat{L}}{x} \in \valtype \Rightarrow \Borrow(\psloc(0,x),\_,\_) \not \in B\\\\
    \hat{S} = \reverse{\outs{\proc}}\sep
    \forall i \in \dom{\hat{S}}.\ \hat{S}[i] \in \mutreftype \Rightarrow \Borrow(\osloc(i),\_,\_) \not \in B
  }
  {\proc,\returnCmd}
  {\st{\hat{L},\hat{S},B}}
  {\st{\hat{L},\hat{S},B}}
  \hspace{3mm}
\pcrule[]
  {
    t = \booleantype
  }
  {\proc,\ao{\branchCmd}{\pc_1,\pc_2}}
  {\st{\hat{L},\stackht{t}{\hat{S}},B}}
  {\st{\hat{L},\hat{S},B}}
  \bigskip

\label{fig:rules-borrow-checker}
\end{centerframe}
\end{figure*}

Having explained the overall structure of our soundness
argument, we now provide intuition for type propagation
on local abstract states.
The rule for operation $\op$ derives a judgment of the form
$
\proc,\op \vdash \st{\hat{L},\hat{S},B} \rightarrow \st{\hat{L}',\hat{S}',B'}
$
if certain conditions are satisfied.
These conditions include availability of appropriately-typed
values in $\hat{L}$ or $\hat{S}$ and absence of certain edges in
the borrow graph $B$.
The state transformation adds or removes a variable-to-type binding in $\hat{L}$,
pushes or pops types in $\hat{S}$, and adds or removes edges in $B$.
The rules for \movelocCmd and \storelocCmd also prevent an input of procedure $\proc$
from being moved or overwritten
to enable accurate tracking of transitive borrow relationships across a procedure call.

The rule for $\ao{\movelocCmd}{x}$ moves the type of variable $x$ to the top of
operand stack.
The rule checks that $x$ is available in $\hat{L}$
and there are no outgoing borrow edges from $\psloc(0,x)$, the abstract
location of $x$, in case $x$ is a value.
The rule also renames the old position of the moved value to its new position.

The rule for $\popCmd$ pops the top of the operand stack and
eliminates the location corresponding to it using a new operation $\mathit{elim}$.
The expression $\elim{B}{u}$ creates a new borrow graph by
eliminating location $u$ in $B$ as follows:
(1)~For every edge $\Borrow(a,p,u)$ coming into $u$ and edge $\Borrow(u,q,b)$
going out of $u$, add an edge $\Borrow(a,\concat{p}{q},b)$.
(2)~Delete all edges coming into and going out of $u$.
The definition of $\elim{B}{u}$ ensures that all transitive borrow relationships
going through the reference at the top of the operand stack are maintained even
when the top is popped.

The two rules for $\ao{\storelocCmd}{x}$ use a combination of the techniques introduced
for handling $\ao{\movelocCmd}{x}$ and $\popCmd$.
If $x$ is available in the local store and is a reference type,
$\psloc(0,x)$ is eliminated in the borrow graph.
If $x$ is available in the local store and is a value type,
then it is checked that there are no borrow edges going out of $\psloc(0,x)$.
In both cases, the position for the previous top of stack is renamed to $\psloc(0,x)$
since it is being moved into variable $x$.

The rule for $\ao{\borrowlocCmd}{x}$ uses the operation $\mathit{factor}$.
The expression $\factor{B}{u}{v}$, where $u$ may but $v$ may not
have borrow edges incident in $B$, creates a new borrow graph by
redirecting edges going out of $u$ to
go out of $v$ and adding a new edge $\Borrow(u,\varepsilon,v)$.
This operation ensures that borrows from $u$ are propagated to $v$.

The first rule for $\ao{\borrowfieldCmd}{f,x}$ addresses the case when
the source reference is mutable and creates a mutable borrow from it.
This rule uses a variation of factor
named $\mathit{factor}_f$, a partial operation with similar inputs as $\mathit{factor}$.
This operation succeeds if there is no edge labeled $\varepsilon$ or $*$ coming out of $u$
in $B$, converting each edge of the form $\Borrow(u,\concat{f}{p},a)$ to $\Borrow(v,p,a)$
and adding the edge $\Borrow(u,f,v)$.
This rule ensures that any borrows from variable $x$ along the field $f$ are
instead borrows from the new reference pushed on the operand stack which is itself
borrowed from $x$.

The second rule for $\ao{\borrowfieldCmd}{f,x}$ addresses the case when
the source reference is immutable and creates an immutable borrow from it.
This rule simply adds a borrow edge labeled $f$ between the source reference
and the new reference.

The rule for $\ao{\copylocCmd}{x}$ is similar to $\ao{\borrowlocCmd}{x}$
in case the variable $x$ being copied is a reference.
The rule for $\writerefCmd$ checks that the target reference does not have
any borrow edges coming out of it.
The rules for $\freezerefCmd$ and $\readrefCmd$ both check that the reference
operand at the top of the stack is {\em freezable}, i.e., all borrowed references
from it are immutable.
If a mutable reference is freezable, it is safe to convert it into an immutable reference.
For space reasons, we skip over the rules for $\packCmd$, $\unpackCmd$,
and $\stackopCmd$
which do not perform any reference-related operations.

The rule for $\ao{\callCmd}{\proc'}$ can be understood as a sequence of simple steps.
First, it checks that no mutable reference being passed to $\proc'$ is borrowed.
Second, it renames the call arguments present on the operand stack to the corresponding
locals in the next frame to simulate the call (see definition of $B_1$).
Third, it simulates the return from the call by adding borrow edges from input
reference parameters to output references returned by the call (see definition of
$B_2$ and $B_3$) and eliminating the locals in the callee frame (see definition of
$B'$).
The definitions of $B_2$ and $B_3$ use the $\mathit{extend}$ operation.
The expression $\extend{B}{\us}{\vs}$, where $\us$ and $\vs$ are sets of locations
with borrow edges incident on locations in $\us$ but no borrow edges incident on
locations in $\vs$, adds an edge $\Borrow(u,*,v)$ for every $u \in \us$ and $v \in \vs$.
Furthermore, the definition of $B'$ uses a generalization of $\mathit{elim}$ that
eliminates a set of locations in the borrow graph one at a time.

The rule for $\returnCmd$ checks that no local of value type is borrowed,
the contents of the operand stack matches the output signature of $\proc$,
and no output that is a mutable reference is borrowed.
This ensures that returned references are valid and the caller's expectations for borrow relationships are sound.

\subsection{Global Memory}
\label{sec:formalizing_global_mem}
We have formalized a borrow analysis that operates on a subset of Move with references to local variables on the call stack. In this section, we informally describe how to amend this model to support global storage. The analysis extensions are straightforward additions to the borrow graph domain that do not require changes to the existing rules.

We model global storage as an extra component $G$ in the concrete program state $\st{P,S,M,G}$, where $G$ maps a type $t \in \valtype$ and an address $a \in \addresstype$ to a location $c \in \cl$.
The extended borrow checker uses the \code{acquires} annotation described in \S \ref{sec:acquires_analysis} to abstract $G$ in the global access instructions described in Section\ref{sec:acquires_analysis}.
We treat each type $t$ in the \code{acquires} list of a procedure $\proc$ as an extra local variable $l_t$ in the intraprocedural borrow checker rules.
The borrow checker ensures that if there is a reference into a value published at $\tup{t,a}$ for any address $a$, then there is a path to this reference from $l_t$ in the borrow graph.
This guarantee is achieved by treating $\ao{\borrowglobalCmd}{t}$ similar to $\ao{\borrowlocCmd}{x}$, which allows us to handle $\ao{\movetoCmd}{t}$ and $\ao{\movefromCmd}{t}$ much like $\ao{\storelocCmd}{x}$ and $\ao{\movelocCmd}{x}$ (respectively).

The borrow checker must perform one additional check: at a $\ao{\movefromCmd}{t}$, $\ao{\borrowglobalCmd}{t}$, or a call to a procedure that has $t$ in its \code{acquires} list, there must be no outgoing edges from $l_t$ in the borrow graph. Additionally, a separate \code{acquires} static analysis checks that a procedure with any such instructions has an \code{acquires} annotation. Together, these ensure that global reference instructions cannot create a dangling reference to global memory.

\section{Implementation}
\label{sec:implementation}

We have implemented two versions of the borrow checking algorithm in Rust: the bytecode analysis described above as part of the Move bytecode verifier (1072 lines) and source code variant in the Move compiler (1807 lines).
The two implementations share a borrow graph abstract domain library (481 lines).
All of these components are open-source\footnote{Withheld for double-blind review}.

\subsection{Borrow Checker in Move Bytecode Verifier}
The bytecode verifier plays the important role of gating the admission of code to the \Libra
blockchain: a module can only be published if it is first certified by the bytecode verifier.
The borrow analysis is a key component of the bytecode verifier,
but it relies on several auxiliary analysis passes that we will briefly describe.
Each pass analyzes a single module in isolation using the type signatures of its dependencies.

\paragraph{Control-flow graph construction}
The bytecode of each procedure is converted into a control-flow graph over a collection of basic blocks.
Each basic block is a non-empty and contiguous sub-sequence of the bytecode such that
control-flow instructions ---\branchCmd or \returnCmd--- only occur as the last instruction
of the basic block.
Together, the basic blocks are a partition of the entire bytecode sequence.
The construction of the control-flow graph attempts to create maximal basic blocks such that there
is no jump into the middle of a block.
Simple checks such as non-empty bytecode and ending with a control-flow instruction are also performed in this analysis.
The granularity of all subsequent analyses is an entire basic block rather than an individual bytecode instruction.

\paragraph{Stack usage analysis}
This analysis ensures that the shape-matching property from \S \ref{sec:abstractSem} holds for all programs.
It tracks the height of the operand stack in each basic block and checks that the heights are equal at each join point.

\paragraph{Value type analysis}
The goal of this analysis is to make sure that a value is used only if has not been moved and that
each bytecode instruction is applied to values of appropriate type.
Each procedure is analyzed separately exploiting the type annotations on inputs and outputs of
called procedures.
This analysis infers unmoved locals and the types of values on the operand stack
using a straightforward dataflow analysis.

\paragraph{Acquires analysis}
This straightforward analysis checks that \code{acquires} annotations on procedures (see \S \ref{sec:formalizing_global_mem}) are correct.

\paragraph{Borrow analysis}
This analysis is the most complex part of reference safety verification.
The borrow checker (\S \ref{sec:abstractSem}) requires that each instruction
be annotated by an abstract state.
However, our analysis computes these annotations using a fixpoint computation based on abstract interpretation.
The fixpoint computation is performed locally for each procedure with the local abstract state as its
abstract domain.
The key new insight enabling this analysis is a suitable join operation for the borrow graph.
This join of $G$ and $H$ is defined operationally as follows:
\begin{enumerate}
  \item Take the union of edges in $G$ and $H$.
  \item For each edge $\Borrow(x,p,y)$ in the result,
  drop the edge if it is subsumed by another edge $\Borrow(x,p',y)$.
\end{enumerate}
It is possible for the resulting graph to have cycles even if $G$ and $H$ are acyclic.
The join fails in this case and an error is reported.
Thus, when a fixpoint is reached successfully,
we are guaranteed that the computed annotations create a well-typed program.

\subsection{Borrow Checker in the Move Compiler}
The bytecode borrow checker is designed to simply and efficiently reject bad code, not help users diagnose reference issues.
However, the Move compiler contains a source-level implementation of the borrow checker that
augments the core algorithm with important tracking information used to provide informative error messages.

In addition, the compiler uses a liveness analysis to improve the programming experience in two ways:
(1) infer whether a use of a source-level variable should emit a $\ao{\copylocCmd}{x}$ bytecode instruction (if $x$ is live) or a $\ao{\movelocCmd}{x}$ instruction (otherwise), and
(2) immediately releasing dead references to prevent errors in the stricter bytecode borrow checker.

\section{Evaluation}
\label{sec:evaluation}

We claim that our analysis is useful, precise, and efficient.

\paragraph{Utility}
\move is a new language that is not (yet) officially supported outside of the
\Libra project.
However, there are two open-source blockchain projects that use \move by maintaining
a fork of the \Libra codebase: dFinance\footnote{\url{https://github.com/dfinance/dvm/tree/master/stdlib/modules}}
and StarCoin\footnote{\url{https://github.com/starcoinorg/starcoin/tree/master/vm/stdlib/modules}}.
Figure \ref{fig:projects} summarizes the use of references in each project.
We observe that references passed across procedure boundaries (\textbf{\&Proc})
and references to global storage (\textbf{GProc}) are common---over half of procedure signatures contain a reference,
and more than a third touch global storage.

Reference-related mistakes (e.g., null dereferences, dangling references) in intricate,
reference-heavy code with mutability are ubiquitous in other languages.
Anecdotally, \move is no different here---we frequently made such mistakes while developing
\textsc{Anon}.
The difference is that \move reports these errors at compile-time with a message that points out
the unsafe action and the borrow that precludes it (see Figure \ref{fig:compiler_error}).
This helps programmers internalize the discipline enforced by the checker and write reference-safe code.

\begin{figure}
\begin{tabular}{l r r | r | r | r}
\textbf{Project} & \textbf{B} & \textbf{Proc} & \textbf{\&Proc} (\%) & \textbf{GProc} (\%) & \textbf{B/ms} \\
\hline
Anon     & 6.9K & 327 & 196 (60) & 151 (46) & 1.2K \\
StarCoin & 6.2K & 351 & 171 (49) & 123 (35) & 1.3K \\
dFinance & 1.2K & 109 &  33 (40) &  30 (28) & 1.3K \\
\hline
\textbf{Total} & 14.6K & 787 & 400 (51) & 304 (39) & 1.2K\\
\end{tabular}

\caption{Usage of references in three \move codebases.
\textbf{B} and \textbf{Proc} quantify the number of bytecodes and declared procedures for each project.
\textbf{\&Proc} lists procedures with a reference in their type signatures and \textbf{GProc} shows procedures that access global storage.
\textbf{B/ms} shows the average number of bytecode instructions analyzed per millisecond on a 2.4 GHz Intel Core i9 laptop with 64GB RAM.}
\label{fig:projects}
\end{figure}

\begin{figure}
\begin{lstlisting}
7| let x = move c;
           ^^^^^^ Invalid $\texttt{move}$ of local 'c'
6| let r = &c.f;
           ---- It is still being borrowed by this reference
\end{lstlisting}

\caption{Error message reported by the \move compiler for the first dangling reference example from \S \ref{sec:dangling_local_refs}.}

\label{fig:compiler_error}
\end{figure}

\paragraph{Precision}
Like any static analysis, the Move borrow checker introduces approximations that may lead it to reject safe programs.
Sources of imprecision include abstracting references either returned by procedures or created on different sides of a
conditional branch, abstracting values in global storage with their types, and preventing returns of global references.

In our experience with Move, we have only encountered expressivity problems with the last restriction.
Issues usually arise when a module \code{M1} wants to give a module \code{M2} the ability to perform arbitrary
writes to a global value of type \code{M1::T}. In these cases, we used workarounds such as exposing
field setters/getters for the global or combining the two modules.

These workarounds are inconvenient but not fatal---a Move module typically encapsulates its global values to enforce key safety invariants, so the pattern of ``sharing'' globals between modules is uncommon. By contrast, eliminating references to global storage altogether would break key programming patterns such as increasing the \code{balance} field of an \code{Account} value in-place without removing it from storage.

\paragraph{Efficiency}
The final column of Figure \ref{fig:projects} quantifies the performance of the analysis using bytecodes analyzed/millisecond as a metric.
The results show that the modular, intraprocedural analysis runs at a consistent rate on projects of different size.
We note that although our current implementation is single-threaded, it would be easy to parallelize analysis of procedures to further increase the speed of the analysis.






\section{Related Work}
\label{sec:related-work}

\paragraph{Borrow-Based Static Analyses}
Rust~\cite{rust} also uses a borrow-based static analysis to prevent dangling references.
Rust's analysis provides reference safety at the source level, whereas Move provides this guarantee
directly for its executable representation via bytecode verification.
The analyses support different language features (e.g., Rust allows references in records, Move
allows mutable references to global state) and require different annotations
(e.g., Rust has reference \emph{lifetime} annotations, Move has procedure \code{acquires} annotations).
We prove that Move's analysis ensures leak freedom (see \S \ref{sec:abstractSem}), but we are not aware of a similar proof for Rust.

There are also differences in the analysis mechanics. There are two descriptions of the Rust borrow checker: one that
abstracts \emph{reference lifetimes} and ensures that the lifetimes of related references are properly
nested \cite{nll} (formalized in \cite{reed2015patina, DBLP:journals/pacmpl/0002JKD18}), and another that
abstracts the relationship between each value and the \emph{set of loans} \cite{polonius_blog_post}
involving the value and prevents accesses to loaned values (formalized in \cite{weiss2019oxide, DBLP:journals/pacmpl/Astrauskas0PS19}).

Move's analysis is philosophically similar to the second approach, but differs by using a borrow graph domain that preserves
structural information about loans and and values. This allows simpler handling of features like
\emph{reborrowing} (the Rust term for creating a copy of a unique reference), which requires Rust formalizations
\cite{weiss2019oxide, DBLP:journals/pacmpl/Astrauskas0PS19} to maintain additional state, but is the same as a normal borrow for
Move. On the other hand, Rust's reference lifetime annotations allow programmers to precisely specify the relationship between
input and output parameters, but Move does not allow this.



\paragraph{Low-Level Enforcement of Memory Safety}

Move ensures reference safety for its executable representation without trusting a compiler using the approach
of the JVM \cite{DBLP:journals/jar/FreundM03, jvm} and CLR \cite{clr}: lightweight analysis run in a bytecode
verifier. Other approaches to certifying low-level memory safety include typed
assembly \cite{DBLP:journals/jfp/MorrisettCGW03}, proof-carrying code
\cite{DBLP:reference/crypt/Necula11}, and \emph{capability machines}, specialized hardware with a memory-safe
instruction set
\cite{DBLP:journals/pacmpl/SkorstengaardDB19, DBLP:journals/toplas/SkorstengaardDB20, DBLP:phd/ethos/Woodruff14}.
Each of these approaches has merits in its targeted application domains. We chose a bytecode language with a
co-designed verifier for Move to accommodate mutable, persistent global storage, gas metering,
resource types\cite{move_white}, and other features required for Move's use-cases.



\bibliography{bibfile}


\begin{thebibliography}{27}


\ifx \showCODEN    \undefined \def \showCODEN     #1{\unskip}     \fi
\ifx \showDOI      \undefined \def \showDOI       #1{#1}\fi
\ifx \showISBNx    \undefined \def \showISBNx     #1{\unskip}     \fi
\ifx \showISBNxiii \undefined \def \showISBNxiii  #1{\unskip}     \fi
\ifx \showISSN     \undefined \def \showISSN      #1{\unskip}     \fi
\ifx \showLCCN     \undefined \def \showLCCN      #1{\unskip}     \fi
\ifx \shownote     \undefined \def \shownote      #1{#1}          \fi
\ifx \showarticletitle \undefined \def \showarticletitle #1{#1}   \fi
\ifx \showURL      \undefined \def \showURL       {\relax}        \fi
\providecommand\bibfield[2]{#2}
\providecommand\bibinfo[2]{#2}
\providecommand\natexlab[1]{#1}
\providecommand\showeprint[2][]{arXiv:#2}

\bibitem[\protect\citeauthoryear{Astrauskas, M{\"{u}}ller, Poli, and
  Summers}{Astrauskas et~al\mbox{.}}{2019}]%
        {DBLP:journals/pacmpl/Astrauskas0PS19}
\bibfield{author}{\bibinfo{person}{Vytautas Astrauskas}, \bibinfo{person}{Peter
  M{\"{u}}ller}, \bibinfo{person}{Federico Poli}, {and}
  \bibinfo{person}{Alexander~J. Summers}.} \bibinfo{year}{2019}\natexlab{}.
\newblock \showarticletitle{Leveraging rust types for modular specification and
  verification}.
\newblock \bibinfo{journal}{\emph{Proc. {ACM} Program. Lang.}}
  \bibinfo{volume}{3}, \bibinfo{number}{{OOPSLA}} (\bibinfo{year}{2019}),
  \bibinfo{pages}{147:1--147:30}.
\newblock


\bibitem[\protect\citeauthoryear{core team}{core team}{2017}]%
        {nll}
\bibfield{author}{\bibinfo{person}{Rust core team}.}
  \bibinfo{year}{2017}\natexlab{}.
\newblock \bibinfo{title}{RFCs: 2094-NLL}.
\newblock
  \bibinfo{howpublished}{\url{https://github.com/rust-lang/rfcs/blob/master/text/2094-nll.md}}.
\newblock


\bibitem[\protect\citeauthoryear{for double-blind review}{for double-blind
  review}{2019a}]%
        {libra_blockchain_white}
\bibfield{author}{\bibinfo{person}{Withheld for double-blind review}.}
  \bibinfo{year}{2019}\natexlab{a}.
\newblock \bibinfo{title}{The Anon {B}lockchain}.
\newblock
\newblock


\bibitem[\protect\citeauthoryear{for double-blind review}{for double-blind
  review}{2019b}]%
        {move_white}
\bibfield{author}{\bibinfo{person}{Withheld for double-blind review}.}
  \bibinfo{year}{2019}\natexlab{b}.
\newblock \bibinfo{title}{Move: A Language With Programmable Resources}.
\newblock
\newblock


\bibitem[\protect\citeauthoryear{for double-blind review}{for double-blind
  review}{2019c}]%
        {libra_consensus_white}
\bibfield{author}{\bibinfo{person}{Withheld for double-blind review}.}
  \bibinfo{year}{2019}\natexlab{c}.
\newblock \bibinfo{title}{State Machine Replication in the Anon {B}lockchain}.
\newblock
\newblock


\bibitem[\protect\citeauthoryear{for double-blind review}{for double-blind
  review}{2020}]%
        {moveprover}
\bibfield{author}{\bibinfo{person}{Withheld for double-blind review}.}
  \bibinfo{year}{2020}\natexlab{}.
\newblock \showarticletitle{The Move Prover}. In
  \bibinfo{booktitle}{\emph{CAV}} \emph{(\bibinfo{series}{Lecture Notes in
  Computer Science}, Vol.~\bibinfo{volume}{12224})},
  \bibfield{editor}{\bibinfo{person}{Shuvendu~K. Lahiri} {and}
  \bibinfo{person}{Chao Wang}} (Eds.). \bibinfo{publisher}{Springer},
  \bibinfo{pages}{137--150}.
\newblock


\bibitem[\protect\citeauthoryear{Freund and Mitchell}{Freund and
  Mitchell}{2003}]%
        {DBLP:journals/jar/FreundM03}
\bibfield{author}{\bibinfo{person}{Stephen~N. Freund} {and}
  \bibinfo{person}{John~C. Mitchell}.} \bibinfo{year}{2003}\natexlab{}.
\newblock \showarticletitle{A Type System for the Java Bytecode Language and
  Verifier}.
\newblock \bibinfo{journal}{\emph{J. Autom. Reason.}} \bibinfo{volume}{30},
  \bibinfo{number}{3-4} (\bibinfo{year}{2003}), \bibinfo{pages}{271--321}.
\newblock


\bibitem[\protect\citeauthoryear{Grech, Kong, Jurisevic, Brent, Scholz, and
  Smaragdakis}{Grech et~al\mbox{.}}{2020}]%
        {DBLP:journals/cacm/GrechKJBSS20}
\bibfield{author}{\bibinfo{person}{Neville Grech}, \bibinfo{person}{Michael
  Kong}, \bibinfo{person}{Anton Jurisevic}, \bibinfo{person}{Lexi Brent},
  \bibinfo{person}{Bernhard Scholz}, {and} \bibinfo{person}{Yannis
  Smaragdakis}.} \bibinfo{year}{2020}\natexlab{}.
\newblock \showarticletitle{MadMax: analyzing the out-of-gas world of smart
  contracts}.
\newblock \bibinfo{journal}{\emph{Commun. {ACM}}} \bibinfo{volume}{63},
  \bibinfo{number}{10} (\bibinfo{year}{2020}), \bibinfo{pages}{87--95}.
\newblock


\bibitem[\protect\citeauthoryear{Jones and Muchnick}{Jones and
  Muchnick}{1979}]%
        {Jones-al:POPL79}
\bibfield{author}{\bibinfo{person}{Neil~D. Jones} {and}
  \bibinfo{person}{Steven~S. Muchnick}.} \bibinfo{year}{1979}\natexlab{}.
\newblock \showarticletitle{Flow Analysis and Optimization of {LISP}-like
  Structures}. In \bibinfo{booktitle}{\emph{POPL}}.
\newblock


\bibitem[\protect\citeauthoryear{Jung, Jourdan, Krebbers, and Dreyer}{Jung
  et~al\mbox{.}}{2018}]%
        {DBLP:journals/pacmpl/0002JKD18}
\bibfield{author}{\bibinfo{person}{Ralf Jung}, \bibinfo{person}{Jacques{-}Henri
  Jourdan}, \bibinfo{person}{Robbert Krebbers}, {and} \bibinfo{person}{Derek
  Dreyer}.} \bibinfo{year}{2018}\natexlab{}.
\newblock \showarticletitle{RustBelt: securing the foundations of the rust
  programming language}.
\newblock \bibinfo{journal}{\emph{{PACMPL}}} \bibinfo{volume}{2},
  \bibinfo{number}{{POPL}} (\bibinfo{year}{2018}),
  \bibinfo{pages}{66:1--66:34}.
\newblock


\bibitem[\protect\citeauthoryear{Lamport}{Lamport}{1984}]%
        {Lamport:1984:UTI:2993.2994}
\bibfield{author}{\bibinfo{person}{Leslie Lamport}.}
  \bibinfo{year}{1984}\natexlab{}.
\newblock \showarticletitle{Using Time Instead of Timeout for Fault-Tolerant
  Distributed Systems.}
\newblock \bibinfo{journal}{\emph{ACM Trans. Program. Lang. Syst.}}
  \bibinfo{volume}{6}, \bibinfo{number}{2} (\bibinfo{date}{April}
  \bibinfo{year}{1984}), \bibinfo{pages}{254--280}.
\newblock
\showISSN{0164-0925}


\bibitem[\protect\citeauthoryear{Lindholm and Yellin}{Lindholm and
  Yellin}{1997}]%
        {jvm}
\bibfield{author}{\bibinfo{person}{Tim Lindholm} {and} \bibinfo{person}{Frank
  Yellin}.} \bibinfo{year}{1997}\natexlab{}.
\newblock \bibinfo{booktitle}{\emph{The {J}ava Virtual Machine Specification}}.
\newblock \bibinfo{publisher}{Addison-Wesley}.
\newblock


\bibitem[\protect\citeauthoryear{Matsakis}{Matsakis}{2018}]%
        {polonius_blog_post}
\bibfield{author}{\bibinfo{person}{Niko Matsakis}.}
  \bibinfo{year}{2018}\natexlab{}.
\newblock \bibinfo{title}{An alias-based formulation of the borrow checker}.
\newblock
  \bibinfo{howpublished}{\url{http://smallcultfollowing.com/babysteps/blog/2018/04/27/an-alias-based-formulation-of-the-borrow-checker/}}.
\newblock


\bibitem[\protect\citeauthoryear{Matsakis and Klock}{Matsakis and
  Klock}{2014}]%
        {rust}
\bibfield{author}{\bibinfo{person}{Nicholas~D. Matsakis} {and}
  \bibinfo{person}{Felix~S. Klock, II}.} \bibinfo{year}{2014}\natexlab{}.
\newblock \showarticletitle{The Rust Language}.
\newblock \bibinfo{journal}{\emph{Ada Lett.}} \bibinfo{volume}{34},
  \bibinfo{number}{3} (\bibinfo{date}{Oct.} \bibinfo{year}{2014}),
  \bibinfo{pages}{103--104}.
\newblock
\showISSN{1094-3641}


\bibitem[\protect\citeauthoryear{Meijer, Wa, and Gough}{Meijer
  et~al\mbox{.}}{2000}]%
        {clr}
\bibfield{author}{\bibinfo{person}{Erik Meijer}, \bibinfo{person}{Redmond Wa},
  {and} \bibinfo{person}{John Gough}.} \bibinfo{year}{2000}\natexlab{}.
\newblock \bibinfo{title}{Technical Overview of the Common Language Runtime}.
\newblock
\newblock


\bibitem[\protect\citeauthoryear{Morrisett, Crary, Glew, and Walker}{Morrisett
  et~al\mbox{.}}{2003}]%
        {DBLP:journals/jfp/MorrisettCGW03}
\bibfield{author}{\bibinfo{person}{J.~Gregory Morrisett}, \bibinfo{person}{Karl
  Crary}, \bibinfo{person}{Neal Glew}, {and} \bibinfo{person}{David Walker}.}
  \bibinfo{year}{2003}\natexlab{}.
\newblock \showarticletitle{Stack-based typed assembly language}.
\newblock \bibinfo{journal}{\emph{J. Funct. Program.}} \bibinfo{volume}{13},
  \bibinfo{number}{5} (\bibinfo{year}{2003}), \bibinfo{pages}{957--959}.
\newblock


\bibitem[\protect\citeauthoryear{Nakamoto}{Nakamoto}{2008}]%
        {nakamoto}
\bibfield{author}{\bibinfo{person}{Satoshi Nakamoto}.}
  \bibinfo{year}{2008}\natexlab{}.
\newblock \showarticletitle{Bitcoin: A peer-to-peer electronic cash system}.
\newblock  (\bibinfo{year}{2008}).
\newblock
\urldef\tempurl%
\url{http://bitcoin.org/bitcoin.pdf}
\showURL{%
\tempurl}


\bibitem[\protect\citeauthoryear{Necula}{Necula}{2011}]%
        {DBLP:reference/crypt/Necula11}
\bibfield{author}{\bibinfo{person}{George~C. Necula}.}
  \bibinfo{year}{2011}\natexlab{}.
\newblock \showarticletitle{Proof-Carrying Code}.
\newblock In \bibinfo{booktitle}{\emph{Encyclopedia of Cryptography and
  Security, 2nd Ed}}, \bibfield{editor}{\bibinfo{person}{Henk C.~A. van
  Tilborg} {and} \bibinfo{person}{Sushil Jajodia}} (Eds.).
  \bibinfo{publisher}{Springer}, \bibinfo{pages}{984--986}.
\newblock


\bibitem[\protect\citeauthoryear{P{\'{e}}rez and Livshits}{P{\'{e}}rez and
  Livshits}{2020}]%
        {DBLP:conf/ndss/0002L20}
\bibfield{author}{\bibinfo{person}{Daniel P{\'{e}}rez} {and}
  \bibinfo{person}{Benjamin Livshits}.} \bibinfo{year}{2020}\natexlab{}.
\newblock \showarticletitle{Broken Metre: Attacking Resource Metering in
  {EVM}}. In \bibinfo{booktitle}{\emph{NDSS}}. \bibinfo{publisher}{The Internet
  Society}.
\newblock


\bibitem[\protect\citeauthoryear{Reed}{Reed}{2015}]%
        {reed2015patina}
\bibfield{author}{\bibinfo{person}{Eric Reed}.}
  \bibinfo{year}{2015}\natexlab{}.
\newblock \bibinfo{booktitle}{\emph{Patina: A formalization of the Rust
  programming language}}.
\newblock \bibinfo{type}{{T}echnical {R}eport}.
  \bibinfo{institution}{University of Washington}.
\newblock


\bibitem[\protect\citeauthoryear{Schneider}{Schneider}{1990}]%
        {Schneider:1990:IFS:98163.98167}
\bibfield{author}{\bibinfo{person}{Fred~B. Schneider}.}
  \bibinfo{year}{1990}\natexlab{}.
\newblock \showarticletitle{Implementing Fault-tolerant Services Using the
  State Machine Approach: A Tutorial}.
\newblock \bibinfo{journal}{\emph{ACM Comput. Surv.}} \bibinfo{volume}{22},
  \bibinfo{number}{4} (\bibinfo{date}{Dec.} \bibinfo{year}{1990}),
  \bibinfo{pages}{299--319}.
\newblock
\showISSN{0360-0300}


\bibitem[\protect\citeauthoryear{Skorstengaard, Devriese, and
  Birkedal}{Skorstengaard et~al\mbox{.}}{2019}]%
        {DBLP:journals/pacmpl/SkorstengaardDB19}
\bibfield{author}{\bibinfo{person}{Lau Skorstengaard},
  \bibinfo{person}{Dominique Devriese}, {and} \bibinfo{person}{Lars Birkedal}.}
  \bibinfo{year}{2019}\natexlab{}.
\newblock \showarticletitle{StkTokens: enforcing well-bracketed control flow
  and stack encapsulation using linear capabilities}.
\newblock \bibinfo{journal}{\emph{Proc. {ACM} Program. Lang.}}
  \bibinfo{volume}{3}, \bibinfo{number}{{POPL}} (\bibinfo{year}{2019}),
  \bibinfo{pages}{19:1--19:28}.
\newblock


\bibitem[\protect\citeauthoryear{Skorstengaard, Devriese, and
  Birkedal}{Skorstengaard et~al\mbox{.}}{2020}]%
        {DBLP:journals/toplas/SkorstengaardDB20}
\bibfield{author}{\bibinfo{person}{Lau Skorstengaard},
  \bibinfo{person}{Dominique Devriese}, {and} \bibinfo{person}{Lars Birkedal}.}
  \bibinfo{year}{2020}\natexlab{}.
\newblock \showarticletitle{Reasoning about a Machine with Local Capabilities:
  Provably Safe Stack and Return Pointer Management}.
\newblock \bibinfo{journal}{\emph{{ACM} Trans. Program. Lang. Syst.}}
  \bibinfo{volume}{42}, \bibinfo{number}{1} (\bibinfo{year}{2020}),
  \bibinfo{pages}{5:1--5:53}.
\newblock


\bibitem[\protect\citeauthoryear{Szabo}{Szabo}{1997}]%
        {szabo_smart_contracts}
\bibfield{author}{\bibinfo{person}{Nick Szabo}.}
  \bibinfo{year}{1997}\natexlab{}.
\newblock \showarticletitle{Formalizing and Securing Relationships on Public
  Networks}.
\newblock \bibinfo{journal}{\emph{First Monday}} \bibinfo{volume}{2},
  \bibinfo{number}{9} (\bibinfo{year}{1997}).
\newblock
\urldef\tempurl%
\url{https://ojphi.org/ojs/index.php/fm/article/view/548}
\showURL{%
\tempurl}


\bibitem[\protect\citeauthoryear{Weiss, Patterson, Matsakis, and Ahmed}{Weiss
  et~al\mbox{.}}{2019}]%
        {weiss2019oxide}
\bibfield{author}{\bibinfo{person}{Aaron Weiss}, \bibinfo{person}{Daniel
  Patterson}, \bibinfo{person}{Nicholas~D. Matsakis}, {and}
  \bibinfo{person}{Amal Ahmed}.} \bibinfo{year}{2019}\natexlab{}.
\newblock \bibinfo{title}{Oxide: The Essence of Rust}.
\newblock
\newblock
\showeprint[arxiv]{1903.00982}~[cs.PL]


\bibitem[\protect\citeauthoryear{Wood}{Wood}{2014}]%
        {ethereum}
\bibfield{author}{\bibinfo{person}{Gavin Wood}.}
  \bibinfo{year}{2014}\natexlab{}.
\newblock \showarticletitle{Ethereum: A secure decentralised generalised
  transaction ledger}.
\newblock  (\bibinfo{year}{2014}).
\newblock
\urldef\tempurl%
\url{https://ethereum.github.io/yellowpaper/paper.pdf}
\showURL{%
\tempurl}


\bibitem[\protect\citeauthoryear{Woodruff}{Woodruff}{2014}]%
        {DBLP:phd/ethos/Woodruff14}
\bibfield{author}{\bibinfo{person}{Jonathan Woodruff}.}
  \bibinfo{year}{2014}\natexlab{}.
\newblock \emph{\bibinfo{title}{{CHERI:} a {RISC} capability machine for
  practical memory safety}}.
\newblock \bibinfo{thesistype}{Ph.D. Dissertation}. \bibinfo{school}{University
  of Cambridge, {UK}}.
\newblock


\end{thebibliography}

\end{document}